\documentclass[fleqn,usenatbib]{mnras}
\usepackage{newtxtext,newtxmath}
\usepackage[T1]{fontenc}
\usepackage{ae,aecompl}
\usepackage{array}

\usepackage{graphicx}	
\usepackage{amsmath}	
\usepackage{amssymb}	

\usepackage{textcomp}

\usepackage{natbib}
\usepackage{amsmath}
\usepackage{graphicx}
\usepackage[colorinlistoftodos]{todonotes}
\usepackage{xcolor}
\usepackage{multirow}
\usepackage{todonotes}
\usepackage{float}
\usepackage{booktabs}
\usepackage{subfig}
\usepackage{lineno}

\title[Machine Learning Classification for variable stars]{Comparing Multi-class, Binary and Hierarchical Machine Learning Classification schemes for variable stars}

\author[Z. Hosenie et al.]{
Zafiirah Hosenie,$^{1}$\thanks{E-mail: zafiirah.hosenie@gmail.com}
Robert Lyon,$^{1}$
Benjamin Stappers,$^{1}$
Arrykrishna Mootoovaloo$^{2}$
\\
$^{1}$Jodrell Bank Centre for Astrophysics, School of Physics and Astronomy, The University of Manchester, Manchester M13 9PL, UK.\\
$^{2}$Imperial Centre for Inference and Cosmology (ICIC), Imperial College, Blackett Laboratory, Prince Consort Road, London SW7 2AZ, UK.\\
}

\date{Accepted 2019 July 15. Received 2019 July 15; in original form 2019 March 06}

\pubyear{2019}

\begin{document}
\label{firstpage}
\pagerange{\pageref{firstpage}--\pageref{lastpage}}
\maketitle

\begin{abstract}
Upcoming synoptic surveys are set to generate an unprecedented amount of data. This requires an automatic framework that can quickly and efficiently provide classification labels for several new object classification challenges. Using data describing 11 types of variable stars from the Catalina Real-Time Transient Surveys (CRTS), we illustrate how to capture the most important information from computed features and describe detailed methods of how to robustly use Information Theory for feature selection and evaluation. We apply three Machine Learning (ML) algorithms and demonstrate how to optimize these classifiers via cross-validation techniques. For the CRTS dataset, we find that the Random Forest (RF) classifier performs best in terms of balanced-accuracy and geometric means. We demonstrate substantially improved classification results by converting the multi-class problem into a binary classification task, achieving a balanced-accuracy rate of $\sim$99 per cent for the classification of $\delta$-Scuti and Anomalous Cepheids (ACEP). Additionally, we describe how classification performance can be improved via converting a `flat-multi-class' problem into a hierarchical taxonomy. We develop a new hierarchical structure and propose a new set of classification features, enabling the accurate identification of subtypes of cepheids, RR Lyrae and eclipsing binary stars in CRTS data.
\end{abstract}

\begin{keywords}
stars: variables- general -- methods: data analysis - Astronomical instrumentation, methods, and techniques.
\end{keywords}



\section{Introduction}

\begin{figure*}
\centering
\includegraphics[width=\textwidth]{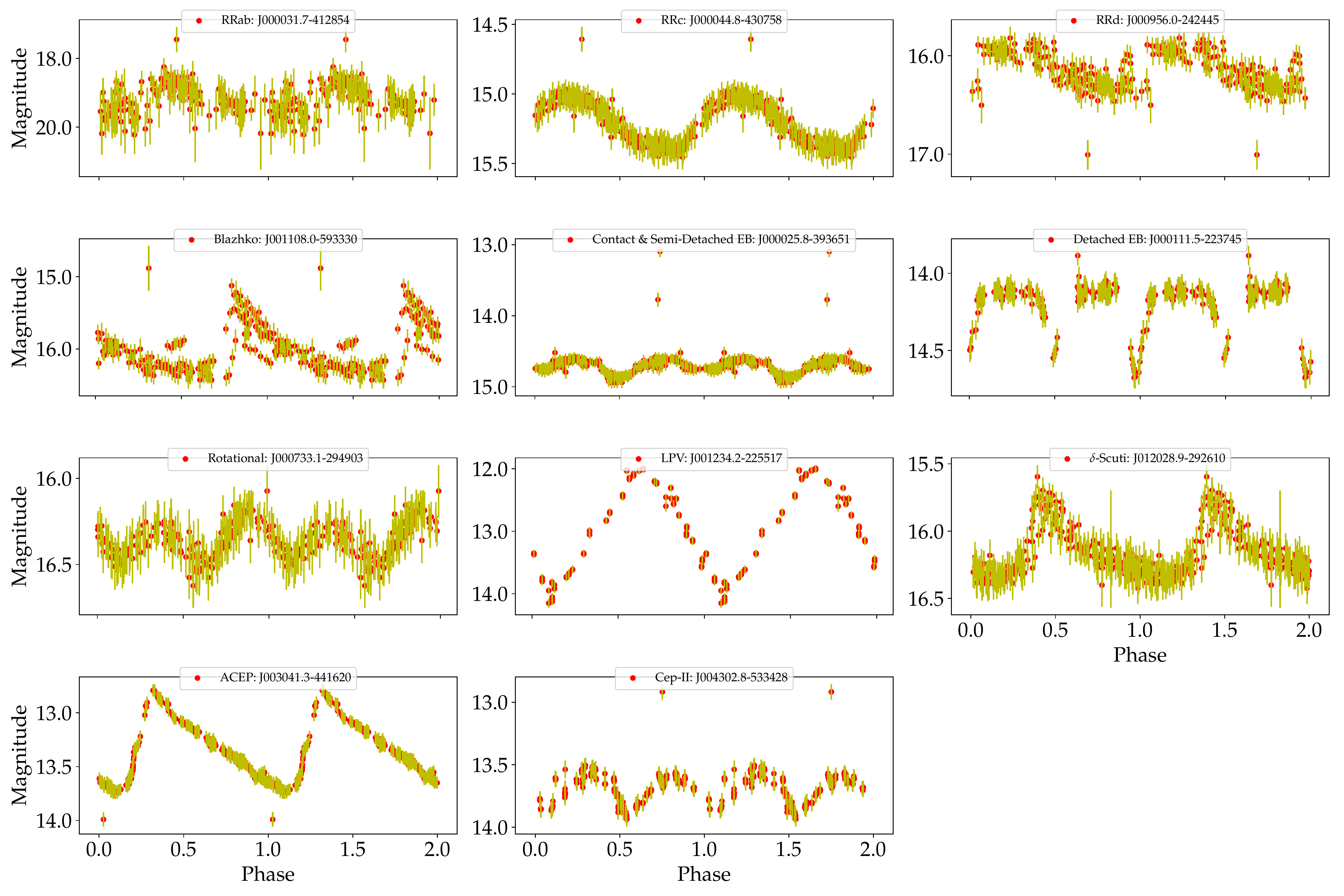}
\caption{\label{fig:Examples of folded-light curves for the various types of variable stars}Examples of folded light curves from the CRTS for the various types of variable stars considered in our analyses.}
\end{figure*}

Astronomy has experienced an increase in the volume, quality and complexity of datasets produced during numerical simulations and  surveys. One factor that contributes to the data avalanche is the new generation of synoptic sky surveys, for example, the Catalina Real-Time Transient Surveys (CRTS) \citep{Drake_2017}. In addition, the Large Synoptic Survey Telescope (LSST, \citet{Ivezic_2008}) for example, which is now on the horizon, will produce $\sim$ 15 Terabytes of raw data per night \citep{Juric_2015}. However, despite this data deluge, source variability is often still visually inspected to detect new promising candidates/variable stars. Visual inspection does have utility for detection and classification. Human experts can extract new useful information despite unevenly sampled data sets and also have the ability to distinguish noisy data from data exhibiting interesting behaviour/characteristics. They can also incorporate complex contextual information into their decision making. However, the efficacy of the manual approach decreases as the volume of data grows exponentially, as will be the case for the next generation of surveys. Visual inspection becomes inconsistent, consequently, mistakes are made, and rare/interesting objects can be missed.

To address this problem, Machine Learning (ML) has been applied to variable-star classification in multiple time-series datasets (see \citealt{Belokurov2003,Willemsen_2007}). In ML, variable stars are represented by features: independent measures that contain information useful for differentiating variable stars into their respective classes. Therefore, several developments have been made towards determining the best methods and features for describing variable-stars, including the Lomb-Scargle periodogram \citep{Lomb_1976,Scargle_1982}, Bayesian Evidence Estimation \citep{Gregory_1992} as well as hybrid methods \citep{Saha_2017}. In addition, \citet{Eyer_2005} analysed the small sharp features of light curves and included them as input features to a Na\"ive Bayes classifier \citep{Zhang_2004}. While \citet{Djorgovski2016} developed an automatic framework to detect and classify transient events and variable stars. They used a subset of the CRTS data to perform classification between two types of variable stars (W Uma and RR Lyrae) and obtained completeness rates of $\sim$96-97 per cent.

\citet{kim2016package} developed the \texttt{UPSILON} package to classify periodic variable stars using 16 extracted features from light curves, which achieves good results. \citet{Mahabal2017} developed a classifier based on the Convolutional Neural Network (CNN) model using labelled datasets of periodic variables from the CRTS \citep{Drakke2009,Djorgovski2011,Mahabal2012,Djorgovski2016}. They transformed a light curve (time series) into a two-dimensional mapping representation ($dm-dt$) which is based on the changes in magnitude ($dm$) over the time-difference ($dt$). Using multi-class classification, their algorithm achieved an accuracy of $\sim$83 per cent. \citet{Narayan_2018} developed an ML approach to classify variable versus transient stars. Similarly, they performed a multi-class classification of combined variable stars \& transients, and a ``purity-driven" sub-categorisation of the transient class using multi-band optical photometry. \citet{Revsbech_2018} used a data augmentation technique to mitigate the effects of bias in their data by generating additional training data using Gaussian Processes (GPs). They used a diffusion map method that calculates a pair-wise distance matrix that outputs diffusion map coefficients of the light curves. These coefficients act as feature inputs to a Random Forest (RF) classifier used to help identifying Type Ia supernova.

We found that it is fundamentally important to develop accurate and robust automated classification methods for this problem using machine learning and other statistical approaches. This paper describes a new automatic classification pipeline for the classification of variable stars via application to archival data.
To our knowledge, this is the first time the southern CRTS \citep{Drake_2017} data set has been used to build/evaluate an automatic classification system.

Similar work has been completed in recent years \citep{kim2016package, Mahabal2017, Narayan_2018}, though the features used for learning are rarely evaluated in a statistically rigorous way. We found that using a large set of features does not imply higher classification metrics. We therefore perform an in-depth analysis of ML features to understand their information content, and determine which give rise to the best classification performance. We utilize various visualization techniques and the tools of Information Theory to achieve this.

Based on our analyses we find that accurate variable star classification is possible with just seven features - much fewer than in other works. In addition, we show that this classification problem cannot be solved with a `flat' multi-class classification approach, as the data is inherently imbalanced. To partially alleviate the `imbalanced learning problem' \citep{Last_2017}, we developed an approach inspired by earlier work in this area \citep{richards2011machine}. This involved converting a standard multi-class problem in to a hierarchical classification problem, by aggregating sub-classes in to super-classes. This results improved performance on rare class examples typically misclassified by multi-class methods. We adopt a similar methodology to \citet{richards2011machine}, however we i) propose a different hierarchical classification structure, ii) use a different feature analysis/selection methodology resulting in different feature choices, iii) apply hyper-parameter optimisation to build optimal classification models, and finally iv) apply the resulting approach to CRTS data.

The outline of this paper is as follows. In \S\ref{sec:data}, we provide a brief description of the dataset used and in \S\ref{sec:Feature-Extraction} we present the feature generation techniques we employ here; while in \S\ref{sec:Classification_Pipeline} we explain how we build the classification pipeline. In \S\ref{Multi-class Classification} we apply state-of-the-art feature visualisation techniques to visualise how separable our features are before performing a multi-class classification. In \S\ref{Binary Classification}, we provide an in-depth feature evaluation to determine the usefulness of our extracted features before performing a binary classification. In \S\ref{sec:Hierarchical_Classification}, we present a hierarchical taxonomy for classification and discuss our results; finally, we summarise our results and conclusions in \S\ref{sec:conclusion}.

\section{DATA}\label{sec:data}

\renewcommand{\arraystretch}{1.75}
\begin{table*}
\begin{minipage}{150mm}

\caption{\label{tab:The-seven-features}The seven features used as inputs to our classification scheme. Six features based on simple statistics, are extracted directly from light curves using FATS. Note that the period feature is obtained from the \citet{Drake_2017} catalog.}
\noindent \begin{centering}
\begin{tabular}{ | c | m{10.0cm}| c | } 
\hline
\textbf{Features} & \textbf{Description} & \textbf{Symbol}\\
\hline
\hline
Mean & $\mu\,=\,\frac{1}{N}\sum_{i=1}^{N}m_{i}$ where $m$ is the magnitude and $N$ is the number of data points. & $\mu$\\
\hline
Standard Deviation & $\sigma=\sqrt{\frac{1}{N-1}\sum_{i=1}^{N}\left(m_{i}-\mu\right)^{2}}$ & $\sigma$\\
\hline 
Skewness & 
$\gamma\,\,=\,\,\frac{N}{\left(N-1\right)\left(N-2\right)}\sum_{i=1}^{N}\left(\frac{m_{i}-\mu}{\sigma}\right)^{3}$  & $\gamma$\\
\hline 
Kurtosis & 
$kurt\,\,=\,\,\frac{N\left(N+1\right)}{\left(N-1\right)\left(N-2\right)\left(N-3\right)}\sum_{i=1}^{N}\left(\frac{m_{i}-\mu}{\sigma}\right)^{4}-\frac{3\left(N-1\right)^{2}}{\left(N-2\right)\left(N-3\right)}$  & $kurt$\\
\hline 
Mean-variance & This is a simple variability index and is defined as the ratio of
the standard deviation, $\sigma$, to the mean magnitude, $\mu$. & $\frac{\sigma}{\mu}$\\
\hline 
Period & \citet{Drake_2017} used the Lomb-Scargle (L-S) periodogram analysis together with an Adaptive Fourier Decomposition (AFD) method to calculate the period of unevenly sampled data. More information about this feature can be found in \citet{Drake_2017}. & $T$\\
\hline 
Amplitude & The amplitude is half of the difference between the
median of the maximum 5\% and the median of the minimum 5\% magnitudes. & $Amp$\\
\hline 
\end{tabular}
\par\end{centering}
\end{minipage}
\end{table*}

We use the publicly available CRTS \citep{Drake_2017} dataset that covers the sky with declinations between $-\,20^{\circ}$ and $-\,75^{\circ}$. The sources in the dataset have median magnitudes in the range 11 < V < 19.5. The dataset contains different forms of periodic variable stars, these are stars that undergo regular changes in brightness every few hours or within a few days or weeks. The periodic variable stars in the dataset can generally be classified into three broad classes: namely eclipsing, pulsating, and rotational. The classes can be further divided into sub-types, for example pulsating stars consist of $\delta$ Scutis, RR Lyrae, Cepheids, and the Long Period Variables (LPVs) group which includes both semi-regular variables and Mira variable stars. The RR Lyrae class consists of RRab\textquoteright s (fundamental mode), RRc\textquoteright s (first overtone mode), and RRd\textquoteright s (multimode). However, many RR Lyrae stars are known to exhibit the Blazhko effect (long-term modulation) \citep{Blazhko_1907}. In addition, the Cepheids include type-II Cepheids (Cep-II), Anomalous Cepheids (ACEPs), and Classical Cepheids. The eclipsing binary variables in the data are divided into detached binaries (EA) and  contact plus semi-detached binaries (Ecl). The rotational class consists of variable stars including the ellipsoidal variables (ELL) and spotted RS Canum Venaticorum (RS CVn) systems. A more detailed overview of the data set is given in \citet{Drake_2017} and \citet{Catelan_2015} gives a more detailed overview of the properties of the various types of pulsating stars. 

\begin{figure}
\centering
\includegraphics[width=0.45\textwidth]{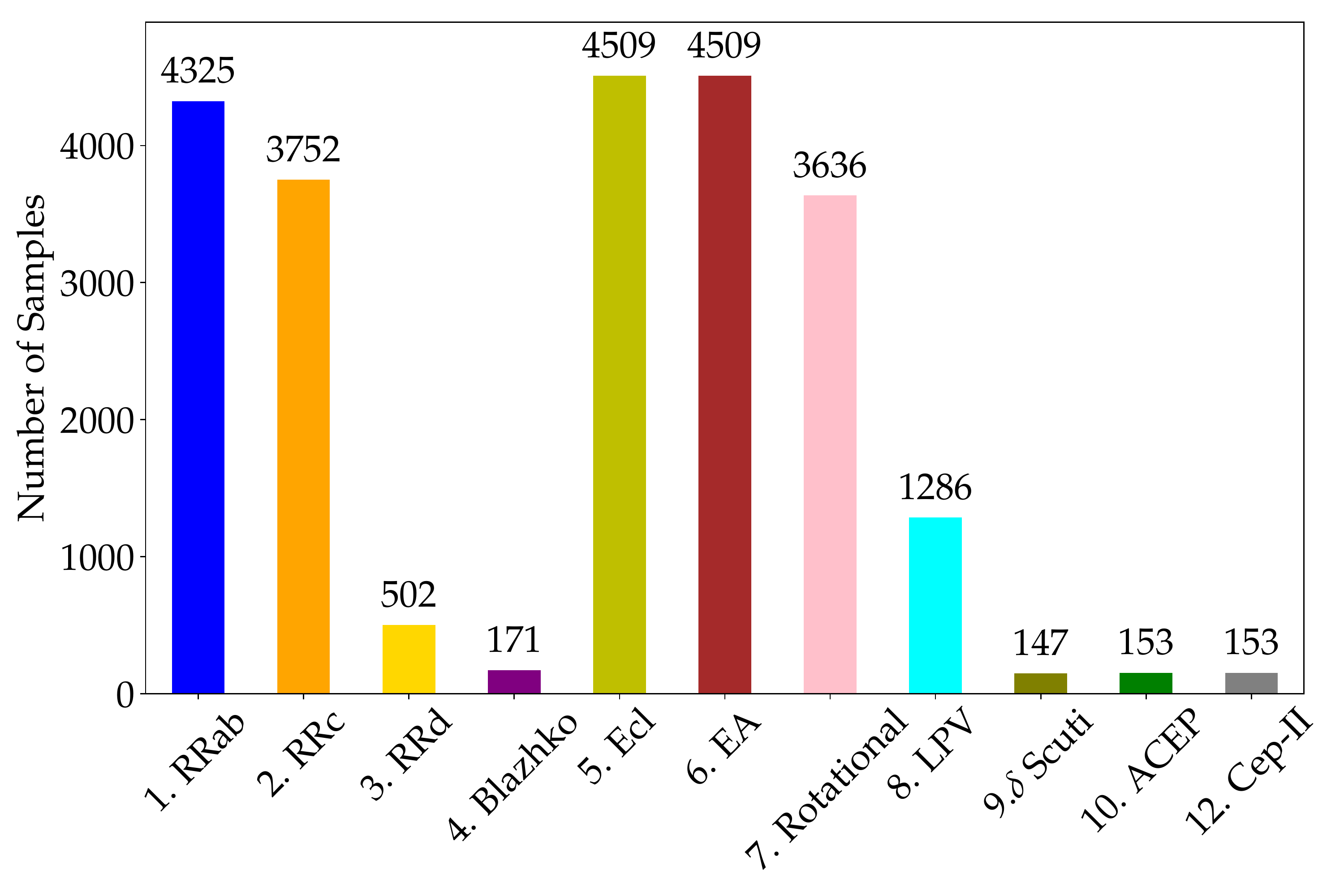}
\caption{\label{fig:Class distributions for the the CSDR2 datasets}Class distributions for the CSDR2 datasets. We downsample Type 5: semi-detached binary stars to 4,509 samples to prevent larger classes from dominating the training sets. The excluded samples $\sim$14,294 for Type 5: Ecl are then included in the test set for prediction. These distributions highlight the remaining imbalance in the datasets.}
\end{figure}

The dataset contains about 37,745 periodic variable stars (Catalina Surveys Data Release 2, \footnote{\href{http://nesssi.cacr.caltech.edu/DataRelease/VarcatS.html}{Catalina Surveys Data Release 2}} CSDR2). For our analysis, we use a sample of 37,437 out of the 37,745 stars from the CSDR2. We have excluded Type 11: Miscellaneous variable stars (periodic stars that were difficult to classify as presented in \citet{Drake_2017}) and Type 13: LMC-Cep-I which as a group consists of only 10 examples. We remove the smallest classes as there are too few samples to characterise those classes, as we also know that classifiers given such data will struggle to categorise them accurately due to the imbalanced learning problem \citep{Last_2017}. Furthermore, we downsample Type 5: semi-detached binary stars to 4,509 samples as this class of object originally consisted of 18,803 samples. We perform this downsampling to prevent the large class from dominating the training sets, which could otherwise potentially bias a classifier. The excluded samples of Type 5 are then included in the test set. Fig. \ref{fig:Examples of folded-light curves for the various types of variable stars} shows examples of folded light curves for each class under consideration.  We also present the number of samples considered for the 11 different types of variable stars in Fig. \ref{fig:Class distributions for the the CSDR2 datasets}. Note that $\sim$14,294 samples of Type 5: Ecl, unused during training, were eventually used in the test set.

\section{Feature Generation}\label{sec:Feature-Extraction}

In general, machine learning algorithms use training data to build a mathematical model, where the training data is comprised of feature data. These features may have varying utility, that is, information rich features are desirable as they can be used to build more accurate classification systems. In this work, we generate statistical features from the light curves to characterize and distinguish different variability classes. Let $S\,=\,\left\{ X_{1},\ldots,\,X_{n}\right\} $, represent the set of variable star data available to us, then $X_{i}$ is an individual star represented by variables known as $features$. An arbitrary feature of star $X_{i}$ is denoted by $X_{i}^{j}$, where $j=1,\,\ldots,\,l.$ Each variable star has an associated label $y$ such that $y\in\,Y=\,\left\{ y_{1,}\ldots,\,y_{k}\right\} $. For multi-class scenarios the possible labels assignable to variable stars vary as $1\leq y\leq12$ but since we do not consider Type 11 examples for reasons given at the end of \S \ref{sec:data}, we note that $y\neq 11$. Meanwhile for binary class classification scenarios, we consider binary labels $y\in\,Y\,=\left[0,\,1\right]$. 

The goal is to build a machine learning algorithm that learns to classify variable stars described by features, from a labelled input vector, also known as the training set, $X_{Train}$. The training set consists of pairs such that $X_{Train}\,=\,\left\{ \left(X_{1},y_{1}\right),\ldots,\,\left(X_{n},y_{n}\right)\right\} $. The learnt mapping function between input feature vectors and labels in $X_{Train}$, can then be utilised to label new unseen stars, in $X_{Test}$. 

In this work, we focus mainly on using the statistical properties of the data with no preconceived notions of their suitability or expressiveness as input features to our ML algorithms. As a result we focus on 7 features, of which 6 are intrinsic statistical properties relating to location (mean magnitude), scale (standard deviation), variability (mean variance), morphology (skew, kurtosis, amplitude), and time (period). These features are highly interpretable, and robust against bias. Note that we remove data points from light curves that are 3$\sigma$ above or below the mean magnitude, where $\sigma$ is the standard deviation and it is an important step to remove any outliers in the data. This cleaning does not alter the light curves significantly as it removes less than 1 per cent of their data points.

Afterwards, we used the FATS\footnote{\href{http://isadoranun.github.io/tsfeat/FeaturesDocumentation.html}{FATS: Feature Analysis for Time Series}} \citep{Nun_2015} Python Library to extract these features. FATS takes as input the unfolded light curves and it outputs various statistical features: the mean, standard deviation, skew, kurtosis, mean-variance, and amplitude. We also incorporate the period for each star given in the catalog as a feature to our ML algorithms. The description of the input features used for classification is listed in Table \ref{tab:The-seven-features}. Practitioners should be cautious when applying the mean magnitude as a feature in combination with data obtained at another telescope. It has the potential to bias a training set against fainter/brighter sources. We note that adding the telescope label as a feature may overcome this issue, but we leave that to future work.

One important aspect of the training process used to build a classification model is data pre-processing. Some ML algorithms, e.g functions/classifiers that calculate the distance between data points, will not work properly without normalisation or standardisation since the range of values of the features/raw data varies widely. We therefore employ a normalisation method, $\hat{\textbf{S}}$,  to standardize the feature data such that all values in the feature space are scaled between 0 and 1. 
 
\begin{figure*}
\centering
\includegraphics[width=0.75\textwidth]{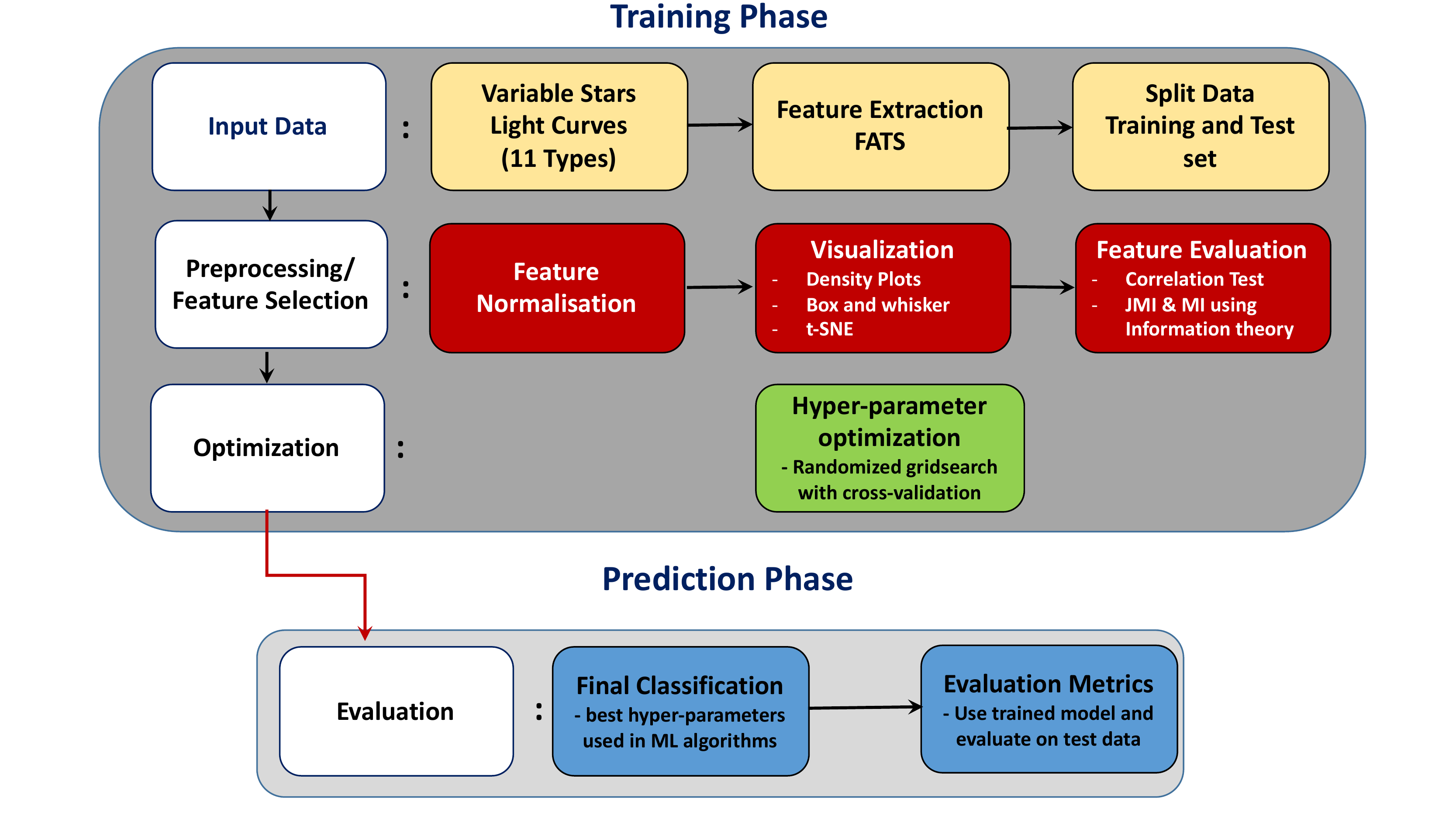}
\caption{\label{fig:Classification_pipeline}The different stages of our classification framework. First, we use light curves of variable stars as input data. For the first stage process, features are extracted using FATS and then are later split into training and test sets. The second stage involves data preprocessing and feature selection. In this process, the extracted features are normalised, visualised and selected based on various techniques. Afterwards, the third stage covers hyper-parameter optimisation using the randomized grid search with cross-validation methods. Finally, the last stage uses the best hyper-parameter to re-train the ML algorithms using the entire normalised training set in stage 2 and evaluate it on the normalised test set in stage 2 using various metrics to quantify the models.}
\end{figure*}

\begin{figure*}
\centering
\includegraphics[width=7cm]{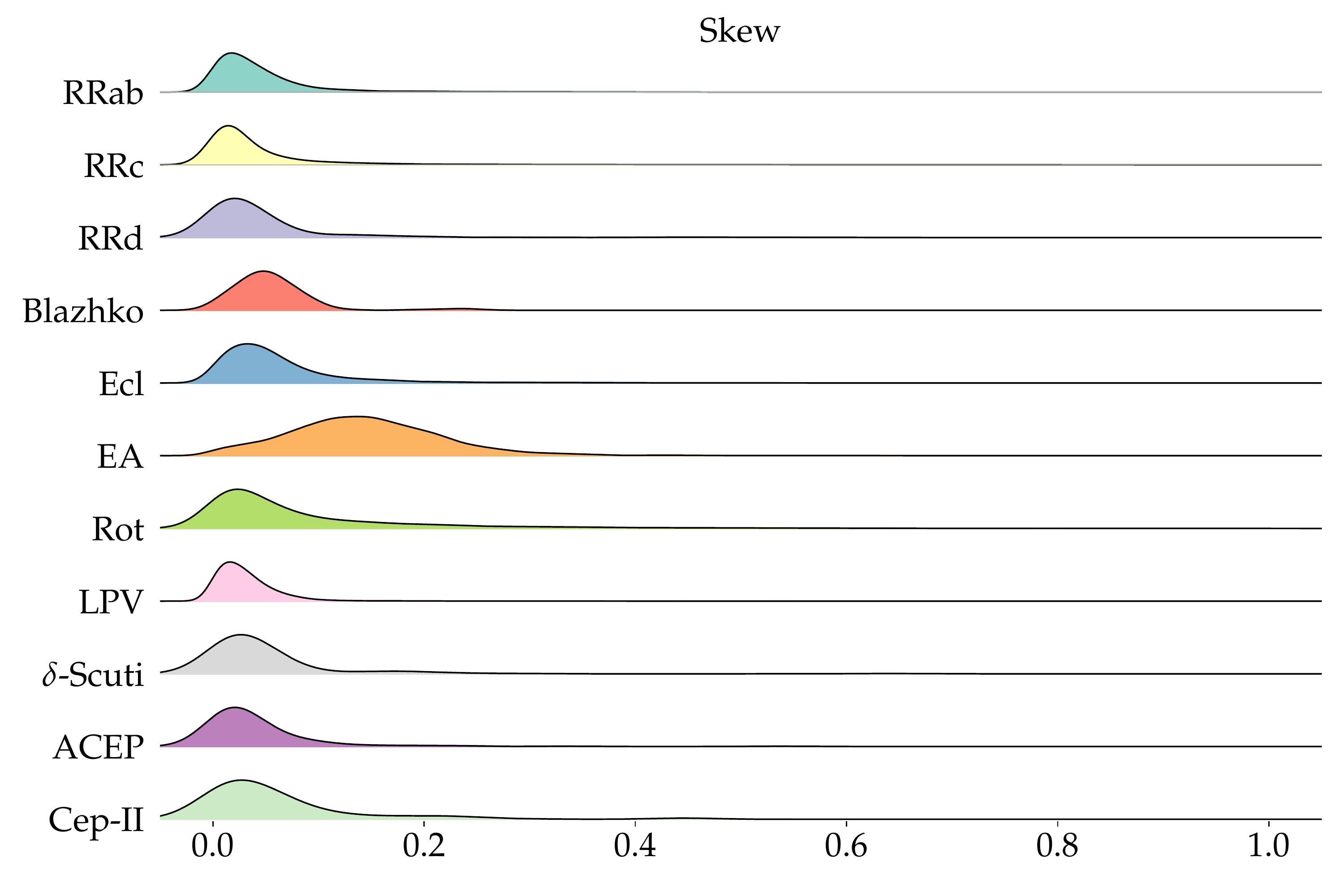} \includegraphics[width=7cm]{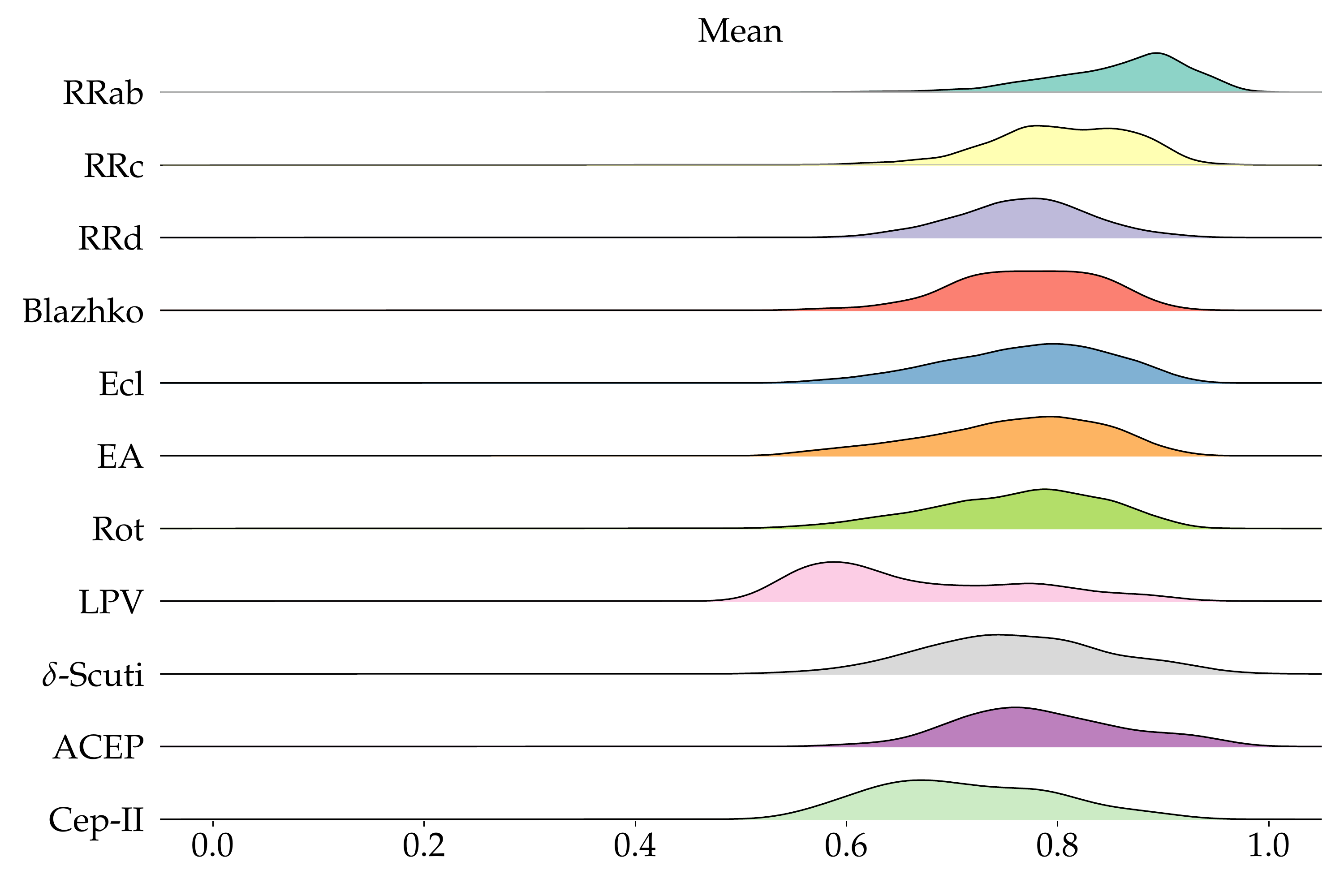}

\centering
\includegraphics[width=7cm]{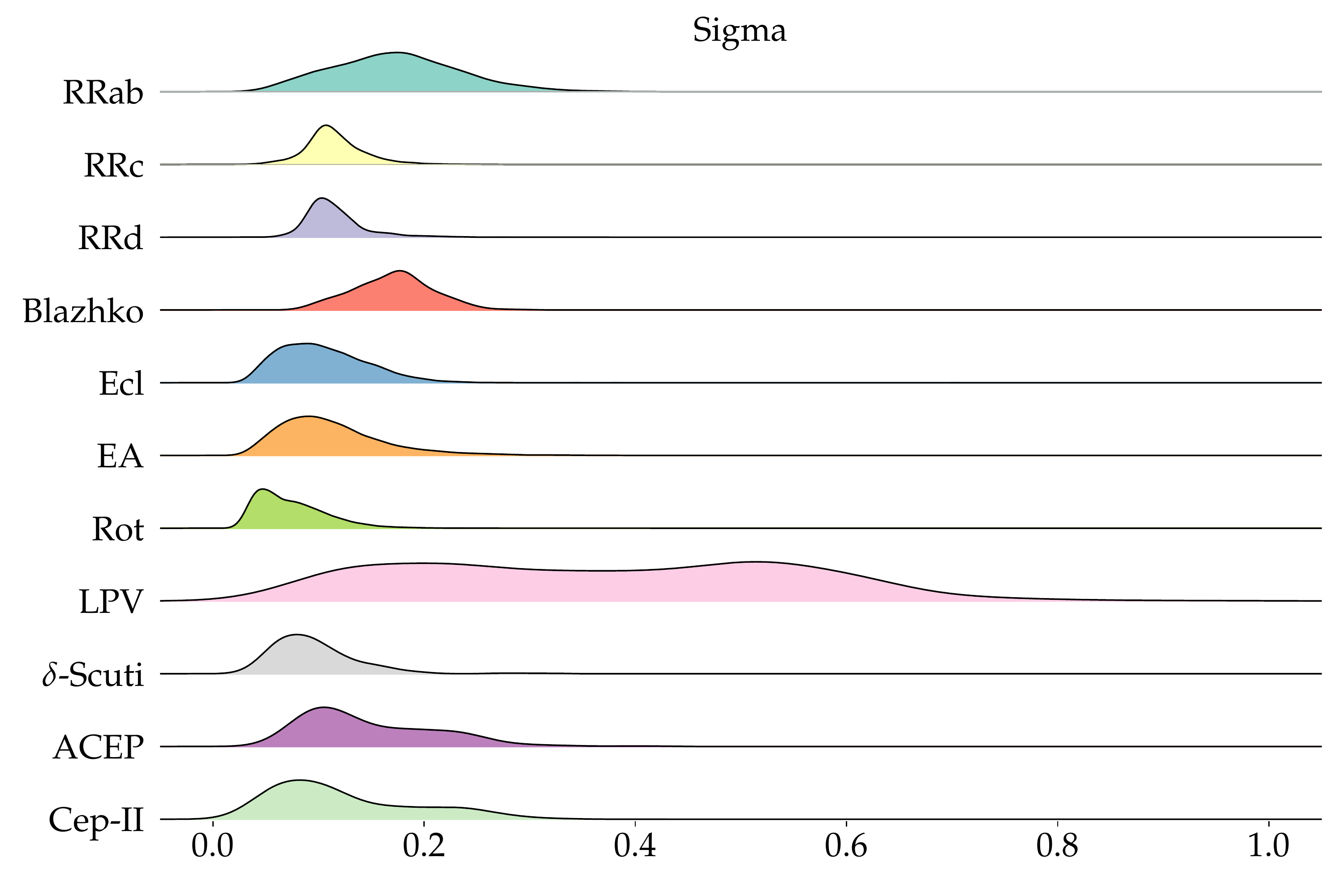} \includegraphics[width=7cm]{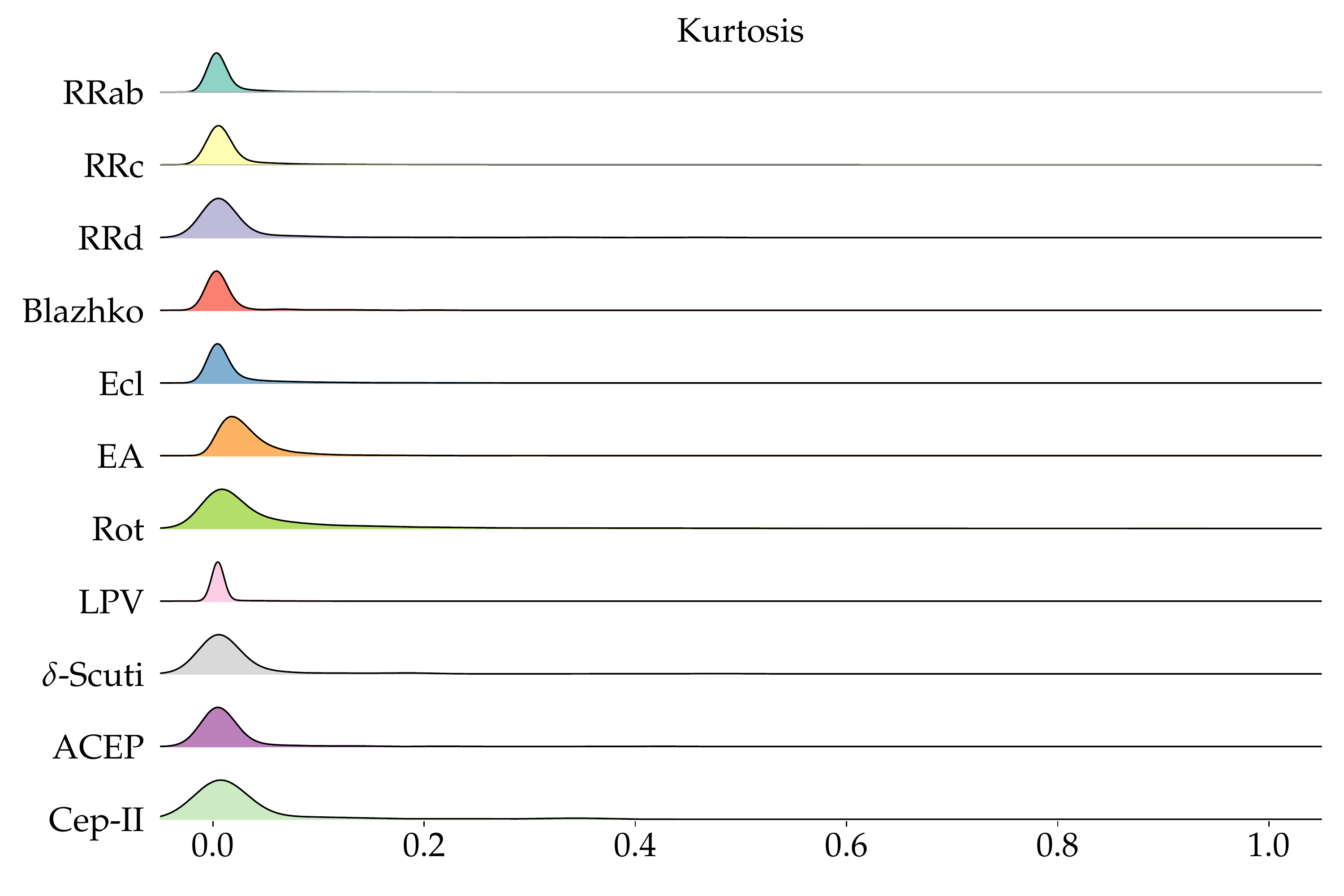}

\centering
\includegraphics[width=7cm]{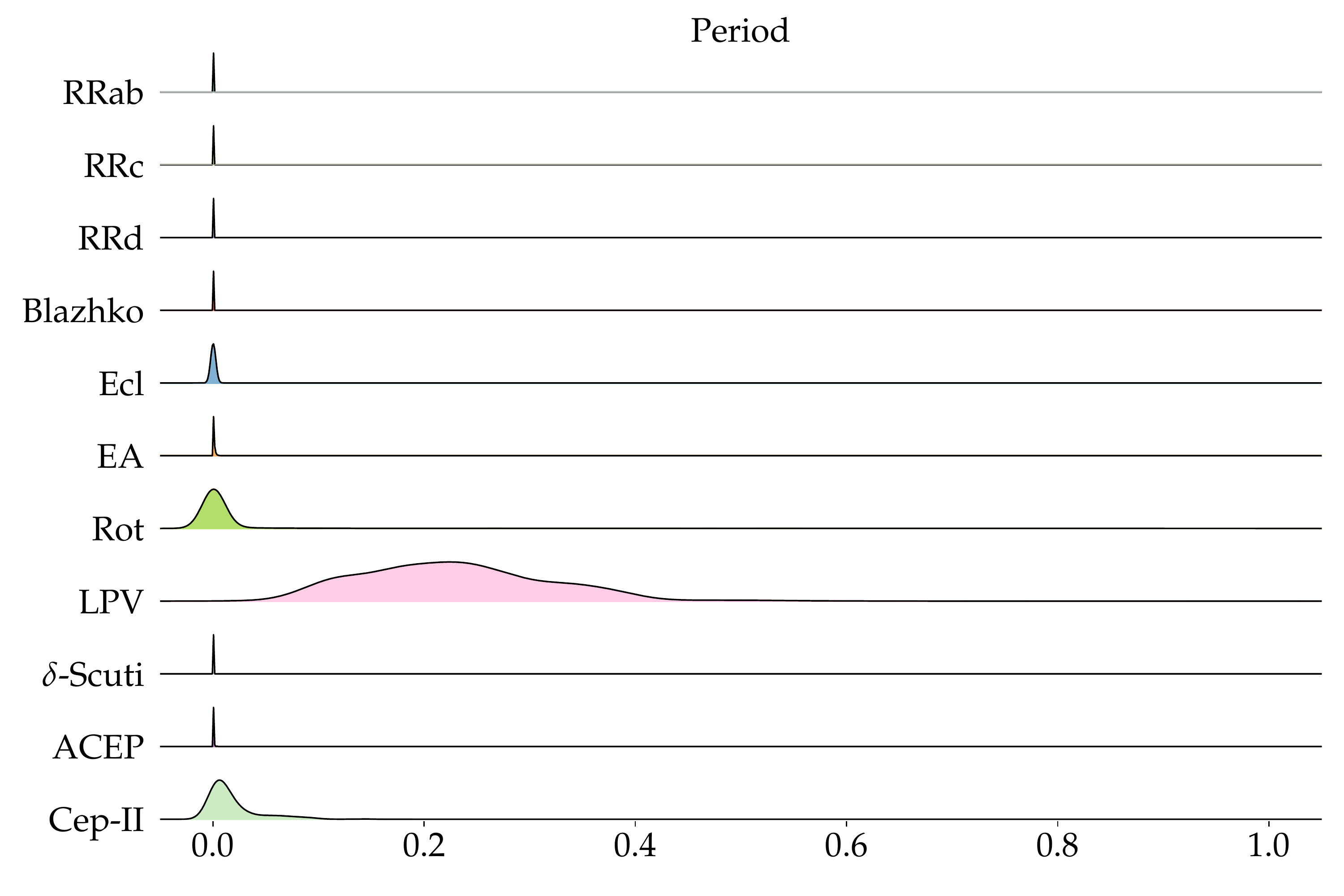} 
\includegraphics[width=7cm]{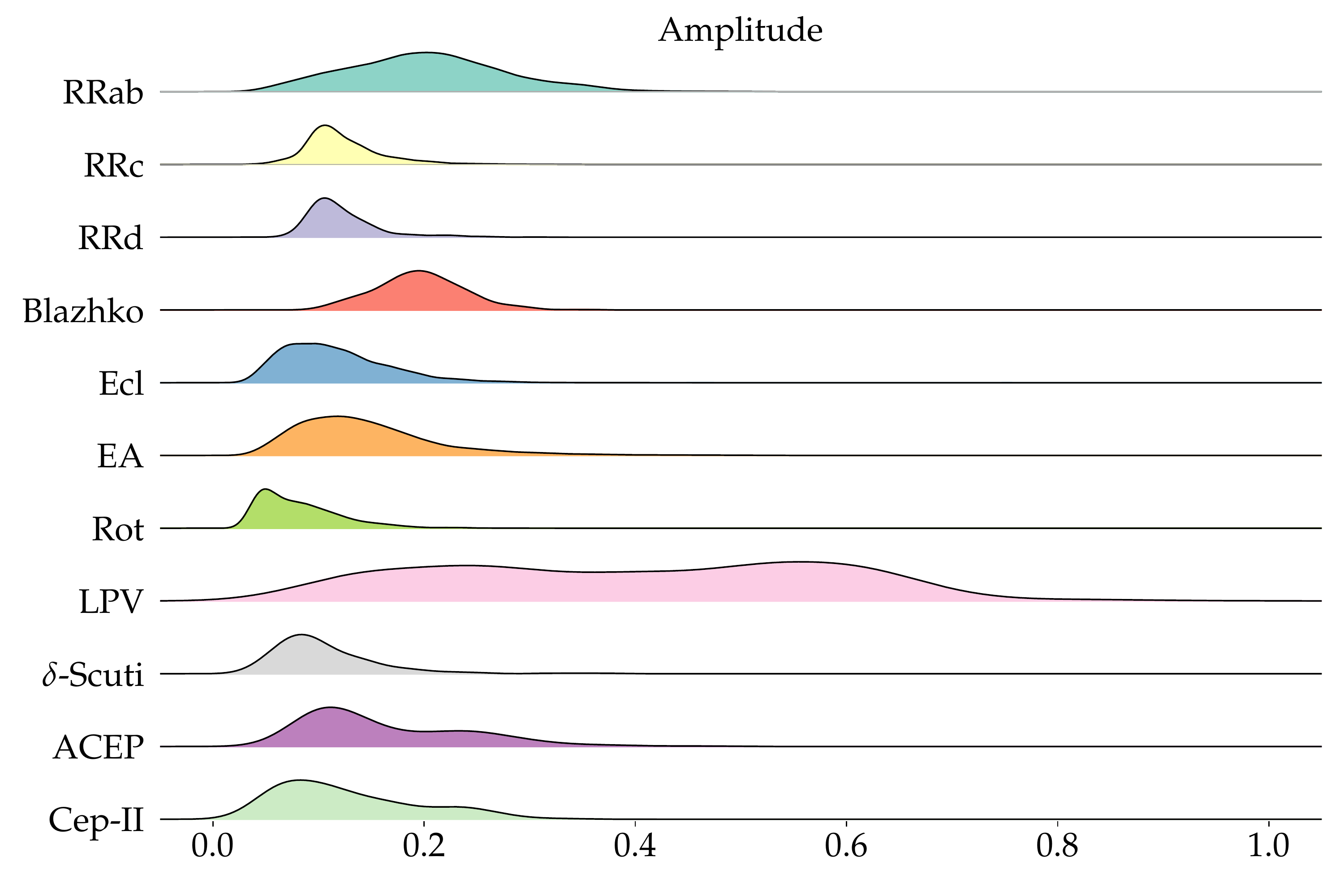}

\centering
\includegraphics[width=7cm]{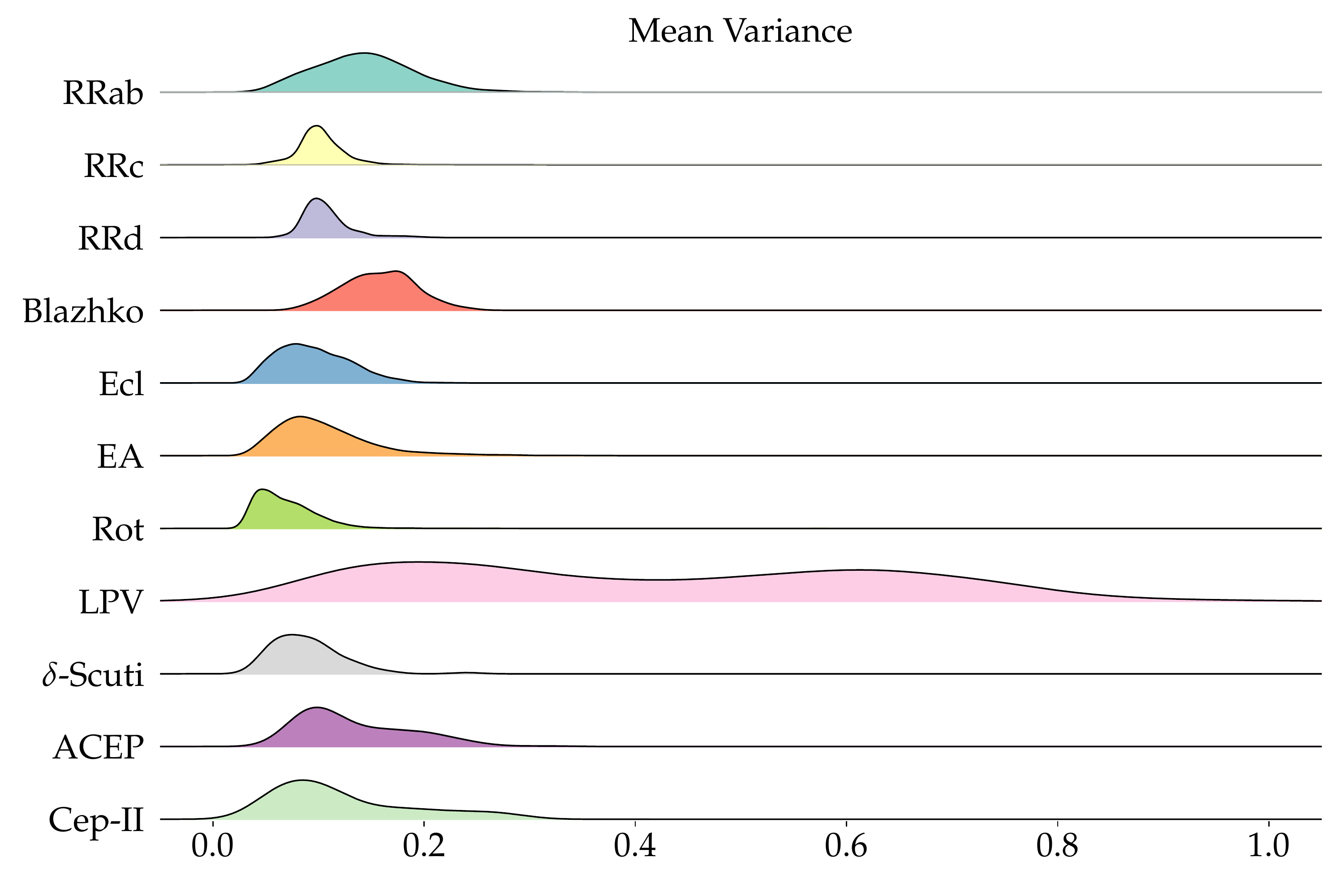}

\caption{\label{fig:One-dimensionality-density-estimates-of-features}One-dimensional density estimates of the selected features by class.
The features considered are (a) Skew, (b) Mean, (c) Sigma, (d) Kurtosis,
(e) Period, (f) Amplitude, (g) Mean Variance.}

\end{figure*}

\section{Classification Pipeline}\label{sec:Classification_Pipeline}

In this section, we describe the process used to perform the classification of variable stars, for instance, training, hyper-parameter optimisation and prediction. We first extract features and then split the feature data into the training (70 per cent) and  the test (30 per cent) data. The training data is used to train the model while the test data is used to evaluate the performance of the trained model. Once the data is split into two, we perform a simple normalization where each feature $X_{i}^{j}$ is divided by its maximum, $max(X_{i}^{j})$ (see Eq. \ref{eq:Normalisation-training-test-set}). The goal of normalisation is to ensure that all features use a common scale. This is beneficial for ML algorithms that are sensitive to feature distributions, for instance, distance-based algorithms that require Euclidean distance computation (for e.g. $k-$Nearest Neighbours ($k$NN)). Some models (e.g. Decision Tree (DT), Random Forest (RF)) are less sensitive to feature scaling. However, it is good practice to standardize when comparing between classification systems to rule out potential sources of disparity in our results. Note that in this context, we found that this simple normalisation was enough to yield good performance. For $X_{Test}$, the features are then rescaled using $max(X_{Train_{i}}^{j})$,

\begin{equation}
X_{Train}=\left|\frac{X_{Train_{i}}}{max\left(X_{Train_{i}}\right)}\right|;\,\,\,\,\,\,\,\,\,X_{Test}=\left|\frac{X_{Test_{i}}}{max\left(X_{Train_{i}}\right)}\right|.\label{eq:Normalisation-training-test-set}
\end{equation}

Afterwards, we apply some feature evaluation strategies to check whether the features show some separability. Then, for our classification purposes, we apply three different classification algorithms: DT, RF and $k$NN, accomplished with Scikit-Learn \citep{Pedregosa}. 
$k$NN \citep{Buturovic_1993} is a simple instance-based technique that assigns an unlabeled example, the label of its $k$ nearest neighbours. This method is based on the Euclidean distance measure. $k$NN performs effectively when the training data set is sufficiently large. However, one disadvantage of $k$NN is that all the features are needed when computing the distance between data points. If a small portion of the data set consists of discriminatory information and the larger portion contains irrelevant features, the distance between the instances will be more influenced by the irrelevant samples and their feature values.

In contrast, Decision trees (DTs) \citep{Quinlan_1986}  attempt to split input data recursively according to feature values. Each split creates a branch, and there can be arbitrarily many branches in a tree. Each branch eventually terminates at a leaf node which is associated with a specific label. The goal of tree learning is to build a tree structure that has decision paths (from tree root to leaf nodes) that accurately separate examples moving down the tree so that they arrive at the correct leaf node (i.e. obtain the correct label). Generally, using a single decision tree for classification often leads to poor performance due to low or high variance. For instance, a small change in the training set can lead to a very different learned tree structure. Given the weakness of individual trees to training variance, multiple trees can be combined to overcome this problem. Any method that combines multiple single-model classifiers in this manner, is known as an ensemble method \citep{Dietterich_2000}. For instance, a Random Forest (RF) \citep{Breiman_2001} is simply an addition of decision trees that aggregate tree decisions, usually leading to improved classification performance. Such ensemble methods have been shown \citep{richards2011machine,Lochner_2016,Narayan_2018} to achieve better results than single-model learners on a variety of datasets.

A major problem faced when using various classifiers is hyper-parameter optimization. This is crucial for finding the hyper-parameters which yield the best overall classification performance for a specific problem. The most widely used techniques are \texttt{Hyperopt} \citep{bergstra2013hyperopt}, a bayesian optimisation approach, grid search and manual search. In our study, we adopt a randomized search that iterates a number of times through pre-specified hyper-parameters and finds the optimum parameter that outputs the best balanced-accuracy for a classifier. Together with the randomized grid search for hyper-parameter optimization, we apply 5-fold stratified cross-validation (see \S \ref{subsec:cross-validation}) on the training set to evaluate model performance, that is, the 70 per cent training set is further split into training and validation sets. We cross-validate to ensure any observed results are real and not just due to some dataset specific effect. We note that CRTS data exhibits distributional disparities - some types of star are common in the data, while others are relatively rare. Traditional machine learning classifiers perform poorly on such data \citep{he2008learning}. They become biased towards correctly classifying the common classes, a strategy that typically yields the greatest overall accuracy. In some cases, this bias can be overcome by re-weighting training examples so that rare examples are weighted higher than common ones. However where/when imbalance is manifested via complex data characteristics (class overlap, small disjuncts, sub-class inseparability, see \citep{he2008learning}), re-weighting alone is insufficient. Based on our analysis of our data presented in \S\ref{subsec:visualisation-1D} \& \S \ref{subsubsec:t-SNE}, we see enough imbalanced characteristics to suggest that our problems cannot be solved by weighting alone, nonetheless we apply re-weighting with the aim of mitigating such problems.

We then proceed with cross-validation, use the best model hyper-parameters found during this process and re-train the model with the entire 70 per cent training set. The trained model is then evaluated on the test set (30 per cent), $X_{Test}$ using various evaluation metrics described in \S \ref{sec:Evaluation_Metrics}. The performance of our pipeline is evaluated on balanced-accuracy, the Geometric-mean score, F1-score, recall and standard confusion matrices. The classification pipeline is summarised in Fig. \ref{fig:Classification_pipeline}.

\begin{figure*}
\centering
\subfloat[t-SNE with large sample data]{\includegraphics[width=0.50\textwidth]{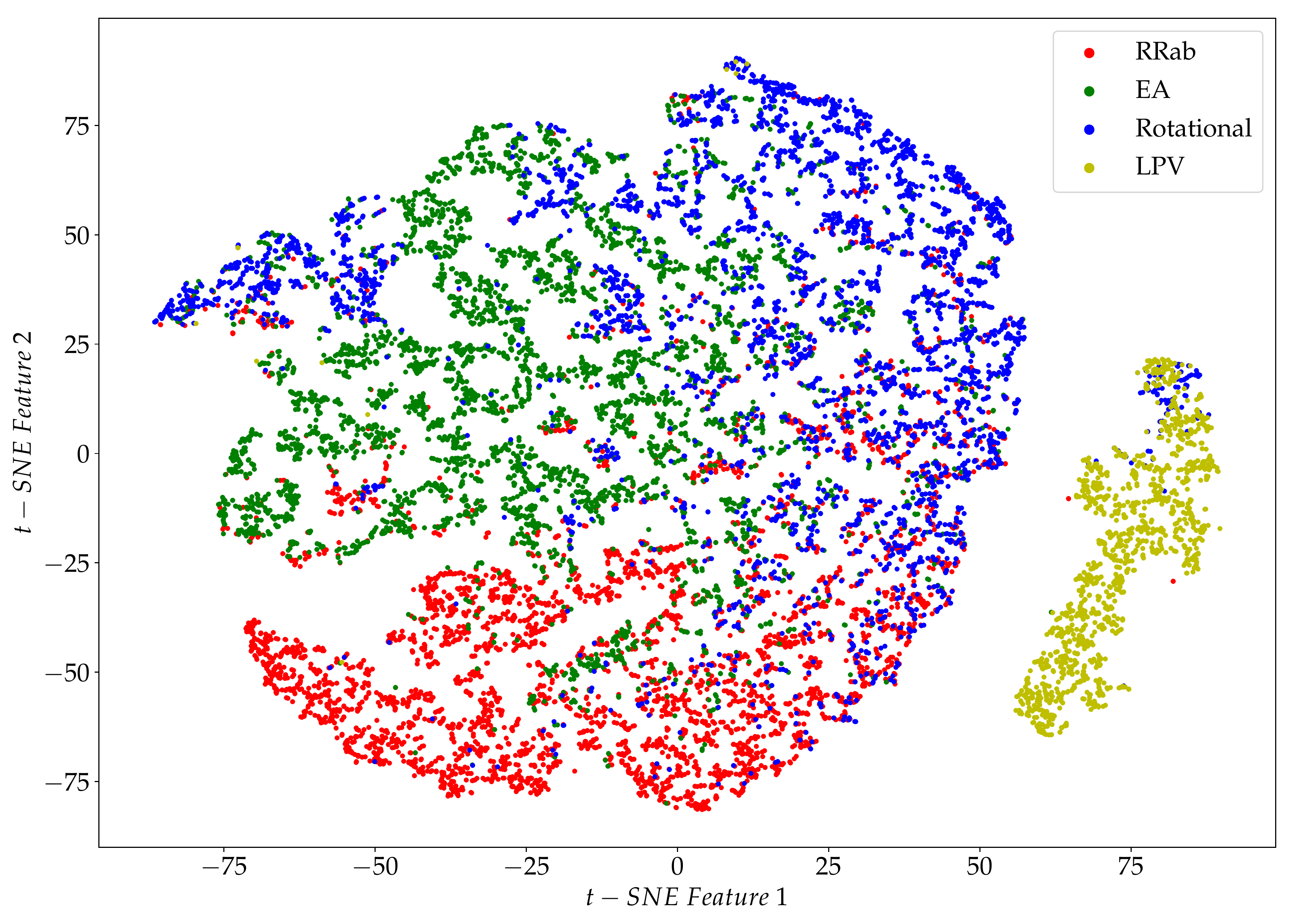}}
\subfloat[t-SNE with small sample data]{\includegraphics[width=0.50\textwidth]{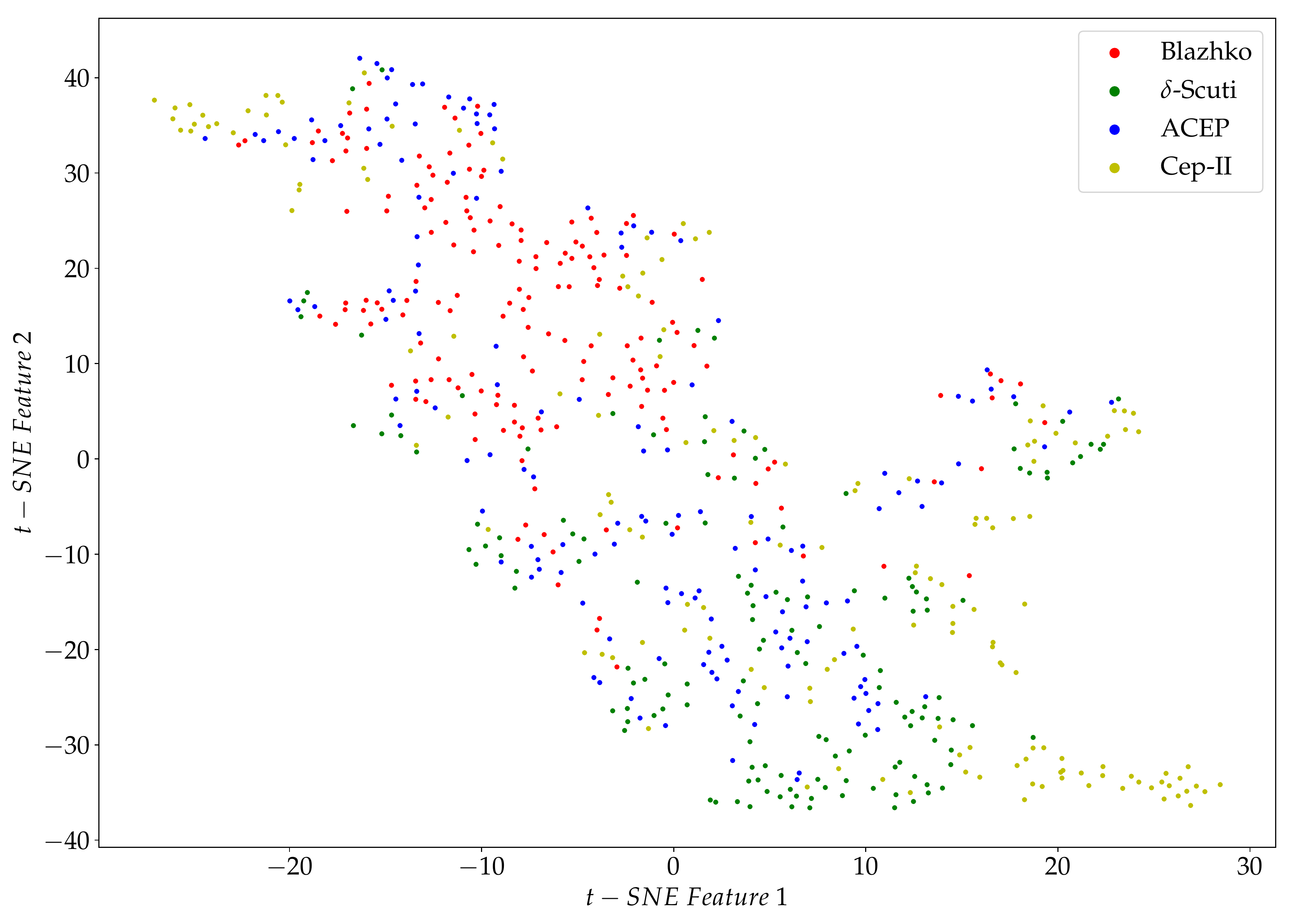}}
\caption{\label{fig:t-SNE-plots}(a) shows the t-distributed stochastic neighbour embedding
(t-SNE) visualization for the feature set after normalisation with large sample data (RRab, EA, Rotational and LPV). We note that the classes are quite well-separated in the embedding space. (b) illustrates t-SNE visualization for the feature set after normalisation with the small sample size data (Blazhko, $\delta$-Scuti, ACEP and Cep-II). No distinct separation is seen within the small sample dataset.}
\end{figure*}

\subsection{$K$-Fold cross validation \label{subsec:cross-validation}}
When using machine learning algorithms, another major problem is an overoptimistic result, i.e the output results are too good to be true on training data, due to over/under fitting. Over-fitting mostly happens when we perform training and evaluation on the same data. Therefore, classification algorithms must be tested on independent data to avoid this problem. One method for avoiding this involves splitting the data into training and testing sets as explained in \S \ref{sec:Feature-Extraction}. Another common method for splitting datasets for evaluation is known as $K$-fold cross-validation, that is, the training dataset $S$ is split randomly into $K$ mutually exclusive subsets $\left(S_{1},S_{2},\ldots,S_{K}\right)$ of nearly the same size. The classification algorithm is trained and tested $K$ times. For each time step $t\in\left\{ 1,2,\ldots,K\right\} $, the algorithm is trained on $K-1$ folds and one fold $S_{t}$ is used for validation. In addition, a stratification of the data is applied such that for each of the $K-folds$, the data are arranged to ensure each fold preserves the percentage of samples for each class in the dataset at large.  The overall balanced-accuracy of an algorithm trained/tested via cross-validation is simply the average of each of the $K$ balanced-accuracy measures obtained after each time step.

\section{Multi-class Classification}\label{Multi-class Classification}
Our main goal is to perform a multi-class classification using our classification pipeline previously described. We first apply some feature evaluation strategies to check whether our extracted features in \S \ref{sec:Feature-Extraction} are good for classification. The most common methods that characterise `good' features: look for the presence of linear correlations between the features and the class labels, and/or we sometimes indirectly measure feature utility by using classification performance (for e.g \cite{Bates_2011a}) as a proxy. If performance is good, we assume the features have utility. However, it is often misleading to evaluate features based on classifier performance as it varies according to the classifier used \citep{brown2012conditional}.

In this work, we employ the three primary considerations as in \citet{Lyon_2016} to evaluate our features. A feature must i) show importance in discriminating between the different classes/types of variable stars, ii) maximise the separation between the various variable stars, and iii) lastly yield a good performance when used in conjunction with a classification algorithm. We have therefore applied two feature visualisation strategies to our features given in Table \ref{tab:The-seven-features} before performing a multi-class classification.

\subsection{Visualisation of feature space with one-dimensional density estimates}\label{subsec:visualisation-1D}

One way to visualise the features we extracted in \S \ref{sec:Feature-Extraction}, is to plot one-dimensional density estimates of the observed distribution as shown in Fig. \ref{fig:One-dimensionality-density-estimates-of-features}. This plot allows us to visually compare the distributions of the normalised features for the different classes of variable stars. The plots depict distinct feature-by-feature characteristics of each class. It enables us to identify outliers and determine if there is multi-modality in the feature space. For example, the RR Lyrae skew distributions are mostly narrow and peaked, showing that these classes are well-characterised by the skewness. In addition, the density plots provide us with information of which features are fundamentally important in discriminating different sets of classes. Some classes have overlapping distributions for one or more feature variables. This applies to the RR Lyrae, eclipsing binary and the Cepheid stars. Whilst other classes, such as the LPVs have trivially separable distributions, suggesting that the LPV class should be easy to classify. However, more rigorous investigation is needed to determine with confidence whether these features are useful for classification purposes.

\begin{figure*}
\centering
\includegraphics[width=0.70\textwidth]{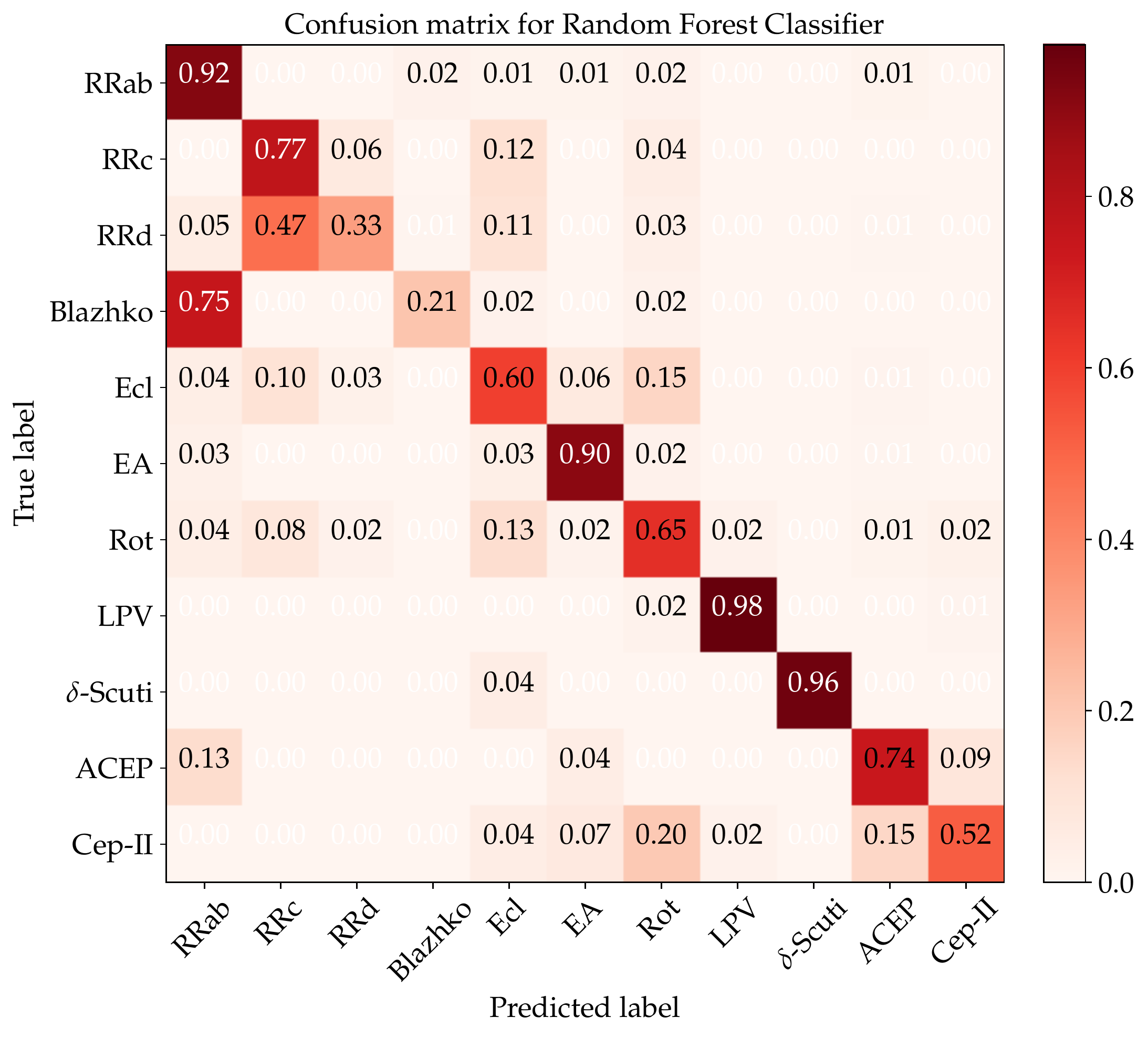}
\caption{\label{fig:Confusion-matrices-for-multi-class-classification}Normalised confusion matrix for multi-class classification with the RF classifier using the best hyper-parameters from a randomized grid search cross-validation. The RF classifier was applied on a 70\% training set and evaluated on a 30\% test set. The classifier performs poorly on under-represented class labels.}
\end{figure*}

\subsection{t-SNE}\label{subsubsec:t-SNE}

After visualising the individual features separately, we wish to understand the relationship between our features but this is difficult to do given the feature dimensionality we are dealing with. Yet it is possible to overcome this problem by reducing the dimensionality so that it is representable in a 2- or 3-D representation space. t-Distributed Stochastic Neighbor Embedding (t-SNE, \citet{vanderMaaten_2008}) is a tool for performing this reduction and visualisation\footnote{\href{http://scikit-learn.org/stable/modules/generated sklearn.manifold.TSNE.html}{sklearn.manifold.TSNE}}.

More specifically, similar objects in high-dimensional space are clustered together with nearby points using a k-D tree \citep{Bentley_1975} of all the points. Using a Student-t distribution (same as the Cauchy distribution \citep{Cauchy_1853}), the Euclidean distance between each point and its $k$-nearest neighbours is computed. This distance is further converted into a probability distribution. Similar points have a high probability of being assigned to the same class and different points have a low probability of being picked. Afterwards, t-SNE constructs a similar probability distribution using a gradient descent method, in the low-dimensional space over the points, thus minimizing the Kullback-Leibler divergence between the two probability distributions \citep{Kullback_1951}.

Generally, classes that are well separated in a t-SNE visualization, yield good levels of classification performance by machine learning systems. However, the converse is not strictly true \citep{Lochner_2016}. We use t-SNE only for the visualization of class separability as there are various limitations with t-SNE, as neither the sizes of the clusters nor distance between the points may be informative \citep{wattenberg2016how}. For our high dimensional visualisation, we partition the data into two categories: i) data with large sample sizes that consists of RRab, RRc, Ecl, EA, Rotational and LPV stars and ii) data with a small sample size containing RRd, Blazhko, $\delta$-scuti, ACEP and Cep-II. Fig. \ref{fig:t-SNE-plots}(a) shows quite well-separated classes for the large sample sizes and we would expect our ML algorithms to perform well using this data. Fig. \ref{fig:t-SNE-plots}(b) strongly suggests inseparability of the classes for the small sample data. This inseparability may be attributed to the fact that there are few examples of each class. Also, another possible explanation is that, for these stars, other features might be required to enable separability. After performing some feature visualisation, we decided to use all of the features as inputs to perform a multi-class classification.

\begin{figure*}
\centering
\includegraphics[width=0.60\textwidth]{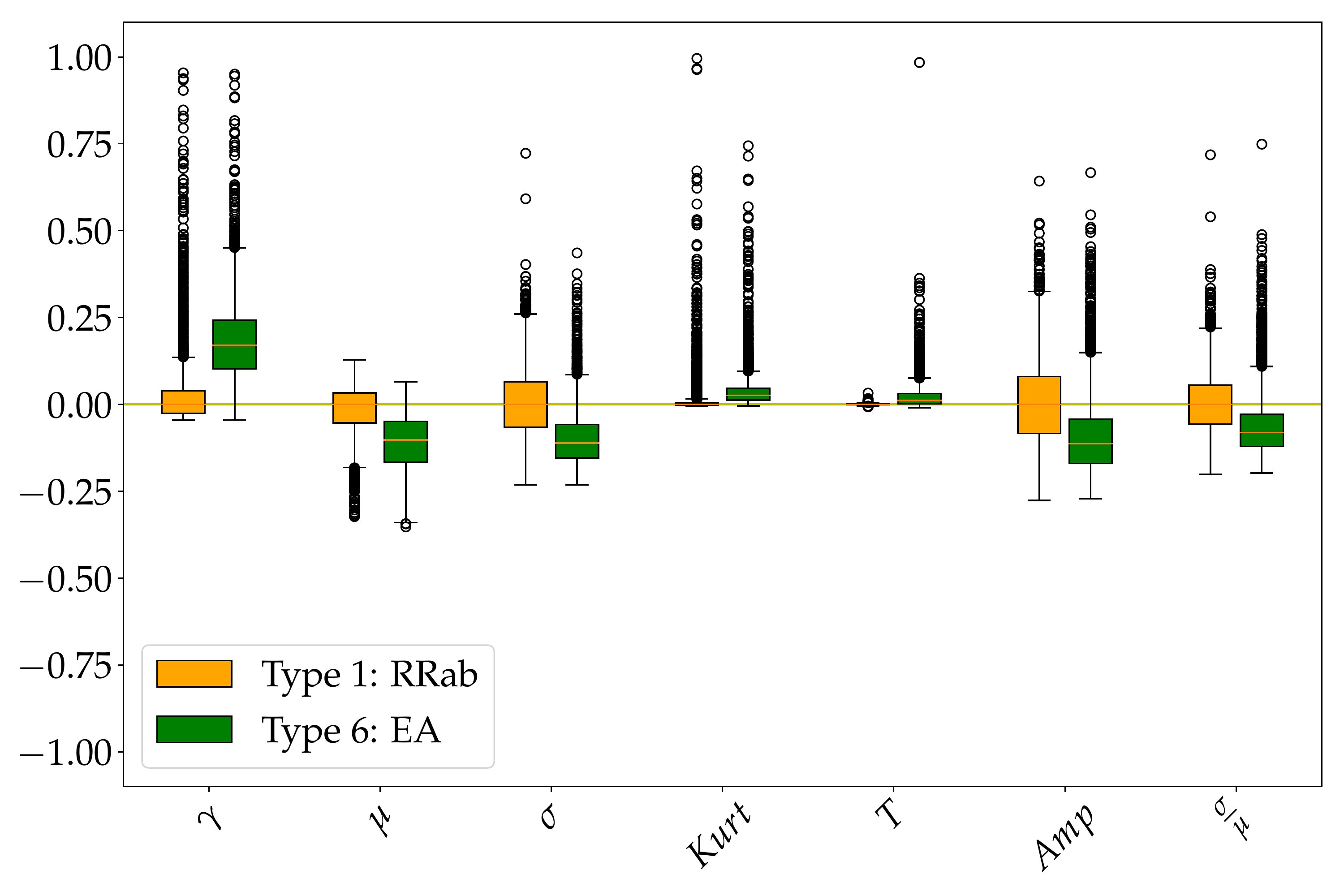}
\caption{\label{fig:BW-Type1-Type6}The box and whisker plots show how the features separate RRab (Type 1) and EA (Type 6) examples. The orange colored boxes describe the feature distribution for RRab sources, where the corresponding black circles represent extreme outliers. The box plots in green describe the EA distributions. There is no clear separation of the period ($T$) between the two classes. The other features show varying degrees of improved separability, and are generally easily separable at a visual level between the two different types. Refer to Table \ref{tab:The-seven-features} for definitions of the features used.}
\end{figure*} 

\subsection{Performance of Multi-class classification}\label{subsec:Multi-class classification}

We implement three classifiers, RFs, DTs and $k$NN to perform an 11-classes classification. Of these, RF achieves the best performance with a  balanced-accuracy rate of $\sim$70 per cent on the 30 per cent test data. This poor performance is not surprising, given the large class imbalance present in the data, i.e. the classes with small sample sizes make up just $\sim$6 per cent of the whole dataset. The results presented here are mainly focused on the RF classifier as this achieves the best performance balanced-accuracy. In Fig. \ref{fig:Confusion-matrices-for-multi-class-classification}, we plot the confusion matrix of the RF classifier which depicts the predicted class versus the true class in a tabular form. The results obtained by a perfect classifier would result in values only along the diagonal of the confusion matrix. The off-diagonals give us information about the various types of errors that the classifier makes. 

We note that some of the mistakes made by the classifier can be attributed to the fact that some of the science classes are physically similar, for example, the RR Lyrae, eclipsing binaries and Cepheids subclasses. We also note that most of the misclassified examples arise from classes for which there are few training examples, hence indicating bias toward larger classes, which illustrates that class imbalance is an issue. Classes comprising of many examples are more likely to be correctly classified, implying that having more examples of the under-represented science classes will help the classification.

\section{Binary Classification}\label{Binary Classification}
Given the complication of the multi-class classification scheme, decomposing the multi-class problem into smaller binary class problems may alleviate this issue. This may reduce the higher complexity inherent to an 11-class problem. Therefore, in this section we consider binary cases for the classification scheme. 

In addition, the poor performance of the multi-class classification in the previous section can be attributed not only to class imbalance  but also to the relative importance of features. Therefore, to further investigate whether the features we used are important for classification, we perform an in-depth analysis of the features used for binary classification in \S \ref{Data visualisation for binary classes}, \ref{Point Biserial Correlation Test} \& \ref{Information Theory} by using roughly balanced classes. Feature selection strategies can be grouped in various categories: classifier-independent (filter methods) and classifier-dependent (wrapper and embedded methods) are the most common. Whilst it is possible to evaluate feature importance using some classification models (e.g. using a random forest), we instead decouple feature analysis from any specific classifier model. We do this as wrapper methods are susceptible to overfitting, which can lead to feature choices that may not generalise well beyond the classifier used during selection \citep{brown2012conditional}. Thus in this paper, we used mostly filter methods. These methods rank the features based on statistical measures of information content, determined by calculating the correlations/relationships between them.

Recall that in  \S \ref{subsubsec:t-SNE}, we sub-divided the data set into two subsets, namely small and large samples, for which each has roughly equal number of examples.  We therefore consider pair-wise combinations of each class within the small-sample dataset and perform feature selection and binary classification. This is repeated for the large-sample dataset. Here, we report on results for Type 1 (RRab) \& Type 6 (EA) only for in-depth feature analyses. However, similar performance is obtained for the other pair-wise combinations. For the performance of binary classification, we report the results obtained with Type 1 (RRab) \& Type 6 (EA) from the large sample dataset and Type 9 ($\delta$-Scuti) \& Type 10 (ACEP) from the small-sample dataset.

\subsection{Data visualisation for binary classes}\label{Data visualisation for binary classes}

The discriminating capabilities of the features are examined for some binary cases. To determine if there is some amount of separability between the two classes, we plot the box and whisker plots for the different features extracted. The data has been pre-processed by performing the normalization described in \S \ref{sec:Feature-Extraction}. Here, we take Type 1 (RRab) as the negative class and Type 6 (EA) as the positive class. For each individual feature, the median value of the negative class is subtracted. This ensures a fair separability scale and centres the plots around the median, hence a distinct separation is seen more clearly between the two classes. The box plots centred on the median value represent the feature distribution of the negative class. Fig. \ref{fig:BW-Type1-Type6} shows a reasonable amount of separability between Type 1: RRab and Type 6: EA. For each individual feature, we note a clear separability between the two classes, except for the Period, $T$. At this point, we are tempted to write-off this feature, however, for a more rigorous analysis of feature selection, we extend our investigation by looking for any linear correlation between the features and the class label.

\subsection{Point Biserial Correlation Test}\label{Point Biserial Correlation Test}

The point biserial correlation coefficient, $r_{pb}$ \citep{Gupta_1960}
is applied to find the relationship between a continuous variable $x$, and binary variable, $y$. Similarly to other correlation coefficients, it varies between -1 and +1, where +1 corresponds to a perfect positive relation and -1 corresponds to a perfect negative relation while a value of 0 means there is no
association at all. The point biserial correlation coefficient is similar to the Pearson product moment \citep{Pearson_1895}. More detailed information can be found in \citet{Lyon_2016}.

\begin{table}
\centering
\caption{\label{tab:The correlation and MI-JMI}The point-biserial correlation coefficient and the Mutual Information $MI(X;Y)$ for each feature for the binary class pair: RRab (Type 1) and EA (Type 6) is illustrated. Higher mutual information is desirable. A high MI value implies that there is a strong correlation between the features and the class labels. In addition, the Joint Mutual Information (JMI) rank is given for each feature. A lower JMI rank is preferred. The JMI ranking illustrates the importance of each feature after taking into account the redundancy between features.}
\begin{tabular}{|m{1.8cm}|c|c|c|}
\hline 
\multirow{3}{*}{Features} & \multicolumn{3}{c|}{Classes}\\
\cline{2-4} 
 & \multicolumn{3}{c|}{Type (1 \& 6)}\\
\cline{2-4} 
 & $r_{pb}$ & MI & JMI\\
\hline 
Skew & 0.648 & 0.503 & 2\\
\hline 
Mean & - 0.547 & 0.260 & 6\\
\hline 
Std & - 0.481 & 0.402 & 4\\
\hline 
Kurtosis & 0.248 & 0.486 & 3\\
\hline 
Period, $T$ & 0.019 & 0.336 & 1\\
\hline 
Amplitude & - 0.375 & 0.307 & 5\\
\hline 
Mean Variance & - 0.368 & 0.174 & 7\\
\hline 
\end{tabular}
\end{table}

Table \ref{tab:The correlation and MI-JMI} illustrates the correlation between the seven features studied and the target class variable, for one binary class. From Table \ref{tab:The correlation and MI-JMI}, it is observed that there are three features that show strong correlations $\left(>\mid0.45\mid\right)$, for instance, the skew, the mean and the standard deviation. It can also be seen from Table \ref{tab:The correlation and MI-JMI} that the $T$ exhibits a weak correlation for this binary class pair. This means that $T$ might not be useful for classification. However, again one should be careful when judging the feature importance based upon their linear correlations. Features having linear correlations close to zero, may have useful non-linear correlations.

\subsection{Information Theory}\label{Information Theory}
To investigate the relative importance of the features, we implement the Information Theoretic method that \citet{Lyon_2016} applied to the pulsar search problem. According to \citet{Guyon_2003}, features that appear to be information poor, may provide new and meaningful information when combined with one or more other features. Information theory is mainly based on the entropy of a feature, $X$. The Mutual Information (MI, \citet{brown2012conditional}), which measures the amount of information between the input feature $X$ and the true class label $Y$ is defined as:

\begin{equation}
I\left(X;Y\right)\,=\,H\left(X\right)\,-\,H\left(X\mid Y\right),\label{eq:Mutual-Information}
\end{equation}

\noindent where $H(X)$ is the entropy and $H\left(X\mid Y\right)$ is the conditional entropy. The conditional probability represents the amount of uncertainty remaining in $X$, after knowing $Y$. Hence, the mutual information is indicative of the amount of uncertainty in $X$ that is removed by knowing $Y$. MI can be zero if and only if $X$ and $Y$ are statistically independent. Therefore, it is useful for features to have high MI. A high MI indicates that a feature is correlated with the target variable, and thus can in principle be used by a classifier to yield accurate classifications.

\begin{table}
\caption{\label{tab:Binary-classification}A summary of the performance results of the various classifiers, ordered from best-performing to worst-performing based on the balanced-accuracy and G-mean for binary classification. Two values are given for all metrics. For each binary case presented, the first values represent metric values for Type 1 (RRab) and Type 9 ($\delta$-Scuti). The second values are the metrics values for Type 6 (EA) and Type 10 (ACEP). In addition, the average of those metrics are summarised in single value.}

\centering
\resizebox{\columnwidth}{!}{%
\begin{tabular}{m{1.2cm} c c c c c }
\hline 
\textbf{Classifiers} & \textbf{Precision} & \textbf{Recall} & \textbf{F1-Score} & \textbf{G-mean} & \textbf{Balanced Accuracy}\tabularnewline
\hline 
\hline 
\multicolumn{6}{c}{\textbf{\textcolor{blue}{Type 1 (RRab) and Type 6 (EA) Classification}}}\tabularnewline
\hline 
\textbf{RF} & 0.97/0.97 & 0.97/0.97 & 0.97/0.97 & 0.97/0.97 & 0.94/0.94\tabularnewline
                 & $\sim$0.97 & $\sim$0.97 & $\sim$0.97 & $\sim$0.97 & $\sim$0.94\tabularnewline
                 
\hline 
\textbf{DT} & 0.96/0.96 & 0.96/0.96 & 0.96/0.96 & 0.96/0.96 & 0.92/0.92\tabularnewline
                   & $\sim$0.96 & $\sim$0.96 & $\sim$0.96 & $\sim$0.96 & $\sim$0.92\tabularnewline
\hline 
\textbf{$K$NN} & 0.95/0.97 & 0.97/0.95 & 0.96/0.96 & 0.96/0.96 & 0.92/0.91\tabularnewline
                        & $\sim$0.96 & $\sim$0.96 & $\sim$0.96 & $\sim$0.96 & $\sim$0.92\tabularnewline

\hline 
\hline
\multicolumn{6}{c}{\textbf{\textcolor{blue}{Type 9 ($\delta$-Scuti) and Type 10 (ACEP) Classification}}}\tabularnewline
\hline 
\textbf{RF} & 1.0/1.0 & 1.0/1.0 & 1.0/1.0 & 1.0/1.0 & 1.0/1.0\tabularnewline
                  & $\sim$1.0 & $\sim$1.0 & $\sim$1.0 & $\sim$1.0 & $\sim$1.0\tabularnewline

\hline
\textbf{DT} & 1.00/0.96 & 0.96/1.00 & 0.98/0.98 & 0.98/0.98 & 0.95/0.96\tabularnewline
                  & $\sim$0.98 & $\sim$0.98 & $\sim$0.98 & $\sim$0.98 & $\sim$0.96\tabularnewline
\hline 
\textbf{$K$NN} & 0.98/0.98 & 0.96/0.98 & 0.97/0.97 & 0.97/0.97 & 0.93/0.94\tabularnewline
                  & $\sim$0.97 & $\sim$0.97 & $\sim$0.97 & $\sim$0.97 & $\sim$0.93\tabularnewline
\hline 

\end{tabular}
}
\end{table}

The MI value of the seven features are listed in Table \ref{tab:The correlation and MI-JMI}. Using the MI method, we are able to choose the features that best describe the difference between two types of classes. However, the selected features might have redundant information, for example two features that give high MI might have the same information, in which a single selected feature could be all that is required to achieve the same classification results. 

Therefore, a ranking system can be applied, which is also known as the Joint Mutual Information (JMI) \citep{Yang_2011} criterion. The JMI enables the detection of complimentary information, and thereby a reduction in the number of features by eliminating redundancy. The JMI performs a selection from a sample of feature sets based on the amount of complementary information and ranks them. It starts from the feature that possesses the largest MI value, $X^{1}$. A greedy iterative process is used to decide which features complement $X^{1}$ \citep{Guyon_2003}.
The JMI score is given by,

\begin{equation}
\textrm{JMI}\left(X^{J}\right)\,=\,\sum_{X^{K}\in F}I\left(X^{J}X^{K};Y\right),
\end{equation}
where $X^{J}X^{K}$ is the joint probability of two features and $F$ is the selected features. The iterative process continues until a set of features are selected or all features are ranked. The use of the JMI enables a reduction in the amount of redundancy within the features. 

We have applied the JMI to our feature data as shown in Table \ref{tab:The correlation and MI-JMI}.
We note that features which appear to be of low information content, as determined via visual analysis and study using the point-biserial correlation coefficient, now appear to be useful. It is tempting to write off features that show low scoring linear correlations. However, it can be seen that even though the Period, $T$, exhibits a low linear correlation, it is ranked as the `$1^{st}$' best feature by the JMI. Given that we have shown that all the seven features have utility, we apply them all for binary classification. 

\subsection{Performance of Binary Classification}\label{subsec:Binary classification}
For binary classification, we train three classifiers independently using roughly balanced pair-wise class combinations from the large-sample and small-sample datasets. We present the results for i) Type 1: RRab \& Type 6: EA and ii) Type 9-$\delta$-Scuti \& Type 10: ACEP only to show the difference between utilizing the two different sized datasets. Similar results are obtained  with other pair-wise combinations of binary classes achieving balanced-accuracy and G-Mean values that vary by $\sim$ $\pm0.03$.

We found that the RF performs best with an overall balanced-accuracy of 0.94 with a G-mean value of 0.97 for Type 1: RRab \& Type 6: EA. In addition, we observe that the two science classes in the small dataset samples (Type 9-$\delta$-Scuti and Type 10: ACEP) have some level of  misclassification in multi-class classification, while in the case of binary classification, the RF (being the best performing algorithm) outputs a balanced-accuracy of 1.0 with a G-mean value of 1.0. The results for both cases studied are summarised in Table \ref{tab:Binary-classification}. 

As discussed in \S \ref{subsec:Multi-class classification}, multi-class classification gave undesirable results. On the other hand, using binary classification (\S \ref{subsec:Binary classification}) we show how to obtain improved balanced-accuracies for the science classes, achieving balanced-accuracies $>90$ per cent in all cases we investigated. We thus now attempt to build upon the result by following an hierarchical approach to classification which will allow us to break the problem into sets of binary comparisons or smaller multi-class comparisons. 

\section{Hierarchical Classification - Results and Discussion}\label{sec:Hierarchical_Classification}
\begin{figure}
\centering
\includegraphics[width=0.45\textwidth]{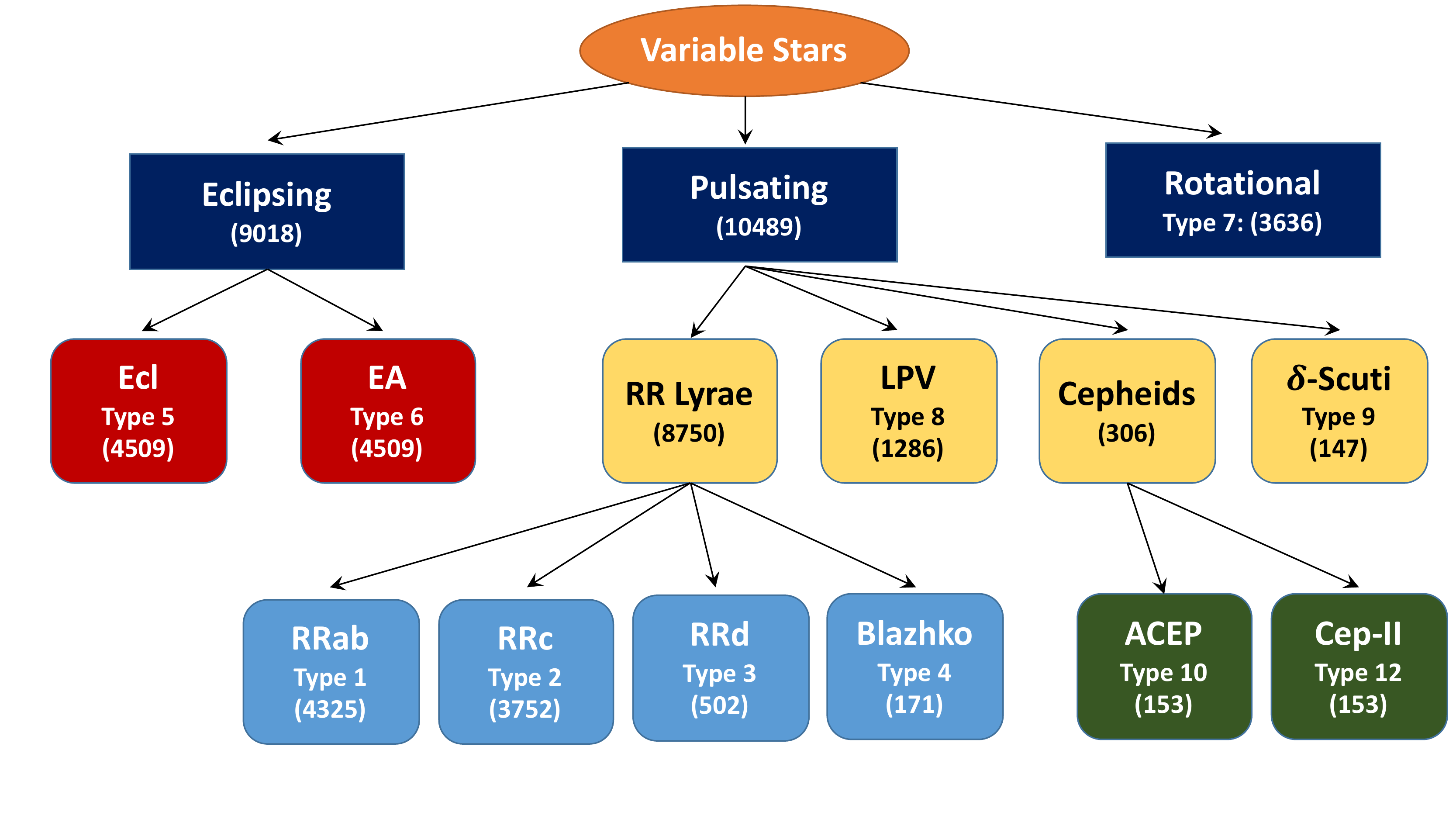}
\caption{\label{fig:A Hierarchical Approach for 11 different types of variable stars}A hierarchy of variable-star classification for data used in \S \ref{sec:data} which is constructed based on the understanding of their physical properties. At the top layer, the variable-stars can be split into three major classes: eclipsing, rotational and pulsating systems. The number in parenthesis represents the number of samples in each particular class.}
\end{figure}

Using some astrophysical properties of the variable stars in our dataset, we group the variable sources into categories as shown in Fig. \ref{fig:A Hierarchical Approach for 11 different types of variable stars}. The first level of the hierarchy is split into three broad science classes: eclipsing, rotational and pulsating. From the top level, we continue to subdivide the classes, for instance, at the third level, we are left with exactly two main classes: RR Lyrae and Cepheids with 6 science sub-classes in the final sub-node. For further details on hierarchical classification, we refer the reader to the excellent review by \citet{Silla_2011}.

Firstly, we perform a multi-class classification on the first hierarchy layer to distinguish between eclipsing, rotational and pulsating stars. This method helps to address some of the class imbalance issues. However, this separation is not perfect because the rotational class has many less examples than the other two classes. The same evaluation methodology as in \S \ref{sec:Classification_Pipeline} is employed and the results for the top layer is summarised in Table \ref{tab:Hierarchical classification}. We obtain a balanced-accuracy rate of $\sim61$ per cent with a G-mean value of $\sim0.79$ for the first layer.

We then move down the hierarchy to perform two separate classifications for layer 2, keeping the same examples in the training sets and the test sets as in the top layer. We subdivide the eclipsing binary group into Ecl and EA classes and perform a binary classification. The result of this experiment shows that we are successful in classifying between the two classes of objects with a 0.86 balanced-accuracy and 0.93 G-mean value. In the second layer, we undertake a multi-class classification of four distinct types of classes: RR Lyrae, LPV, Cepheids and $\delta$-Scuti. We achieve a balanced-accuracy of 0.98 in distinguishing between those classes.

\begin{table*}

\centering
\caption{\label{tab:Hierarchical classification}A summary of the performance results of the RF classifier for the variable-star hierarchical classification. We report metrics per class, separated by `/'. Also, we compute the average metrics taking into consideration the overall classes, summarised in a single value.}

\begin{tabular}{m{3.1cm} m{3.1cm}  m{3.1cm} m{3.1cm} m{3.1cm}}
\hline 
\textbf{Precision} & \textbf{Recall} & \textbf{F1-Score} & \textbf{G-Mean} & \textbf{Balanced Accuracy}\tabularnewline
\hline 
\hline 
\multicolumn{5}{c}{\textbf{\textcolor{blue}{First Level: Eclipsing, Rotational and Pulsating Classification}}}\tabularnewline

0.97/0.19/0.51 & 0.66/0.74/0.87 & 0.79/0.30/0.64 & 0.78/0.78/0.86 & 0.59/0.60/0.75\tabularnewline
$\sim$0.86 & $\sim$0.70 & $\sim$0.74 & $\sim$0.79 & $\sim$0.61\tabularnewline
\hline 
\hline 
\multicolumn{5}{c}{\textbf{\textcolor{blue}{Second Level: RR Lyrae, LPV, Cepheids and $\delta$-Scuti}}}\tabularnewline

1.00/1.00/0.75/0.96 & 0.99/0.99/0.95/1.00 & 0.99/1.00/0.84/0.98 & 0.99/1.00/0.97/1.00 & 0.98/0.99/0.93/1.00\tabularnewline
$\sim$0.99 & $\sim$0.99 & $\sim$0.99 & $\sim$0.99& $\sim$0.98 \tabularnewline
\hline
\multicolumn{5}{c}{\textbf{\textcolor{blue}{Second Level: Ecl and EA}}}\tabularnewline

0.90/0.50 & 0.92/0.94 & 0.95/0.65 & 0.93/0.93 & 0.86/0.86 \tabularnewline
$\sim$0.95 & $\sim$0.92 & $\sim$0.93 & $\sim$0.93 & $\sim$0.86\tabularnewline
\hline
\hline 
\multicolumn{5}{c}{\textbf{\textcolor{blue}{Third Level: RRab, RRc, RRd and Blazhko}}}\tabularnewline

0.96/0.93/0.36/0.29 & 0.98/0.90/0.44/0.19 & 0.97/0.91/0.40/0.23 & 0.97/0.92/0.65/0.44 & 0.94/0.85/0.40/0.18\tabularnewline
$\sim$0.90 & $\sim$0.90 & $\sim$0.90 & $\sim$0.92 & $\sim$0.86\tabularnewline

\hline

\multicolumn{5}{c}{\textbf{\textcolor{blue}{Third Level: ACEP and Cep-II}}}\tabularnewline

0.86/0.95 & 0.96/0.85 & 0.91/0.90 & 0.90/0.90 & 0.82/0.80\tabularnewline
$\sim$0.91 & $\sim$0.90 & $\sim$0.90 & $\sim$0.90 & $\sim$0.81\tabularnewline
\hline
\end{tabular}

\end{table*}

\begin{figure*}
\centering
\subfloat[First level]{\includegraphics[width=0.45\textwidth]{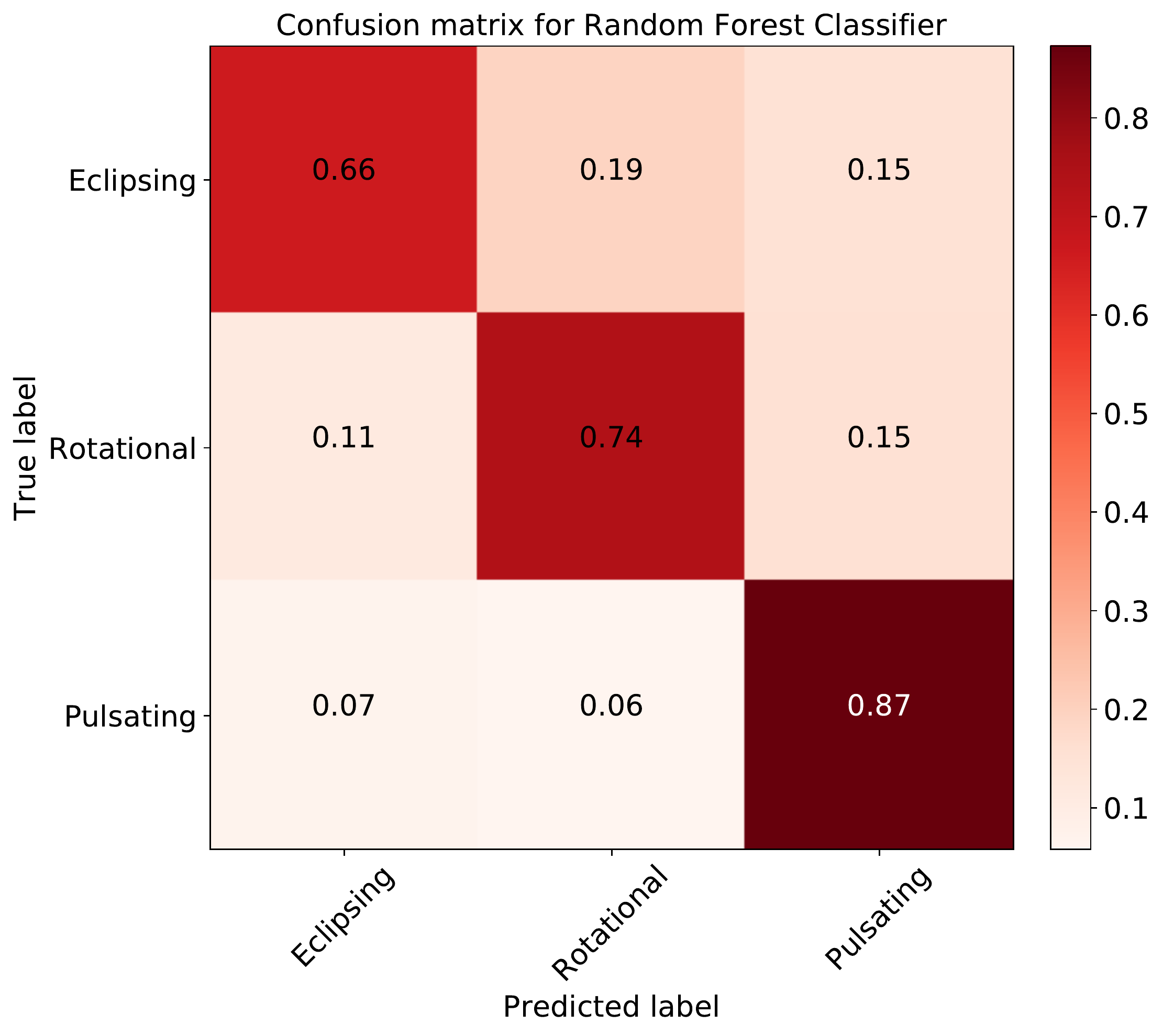}}\\
\subfloat[Second level]{\includegraphics[width=0.45\textwidth]{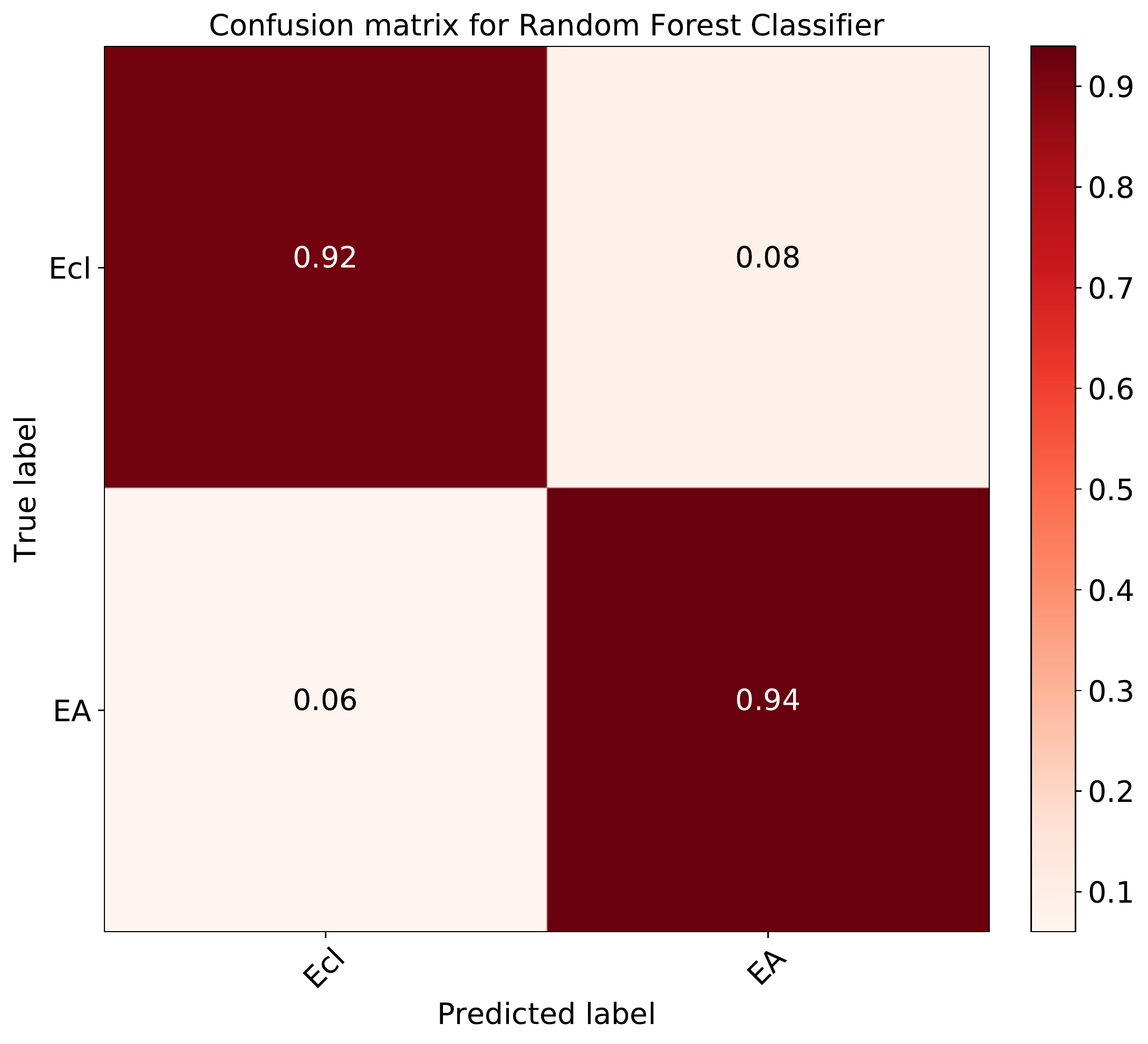}}
\subfloat[Second level]{\includegraphics[width=0.45\textwidth]{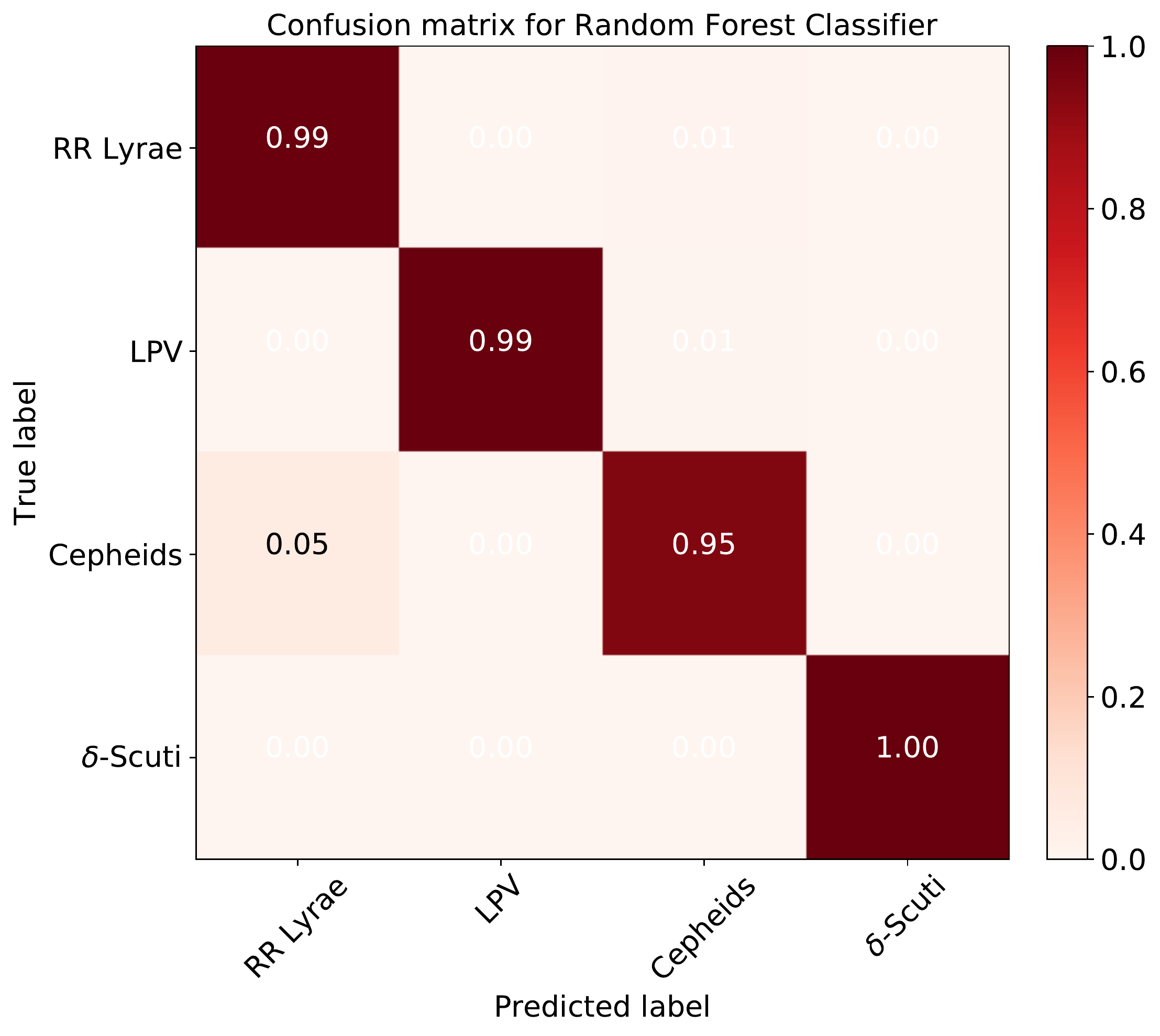}}\\
\subfloat[Third level]{\includegraphics[width=0.45\textwidth]{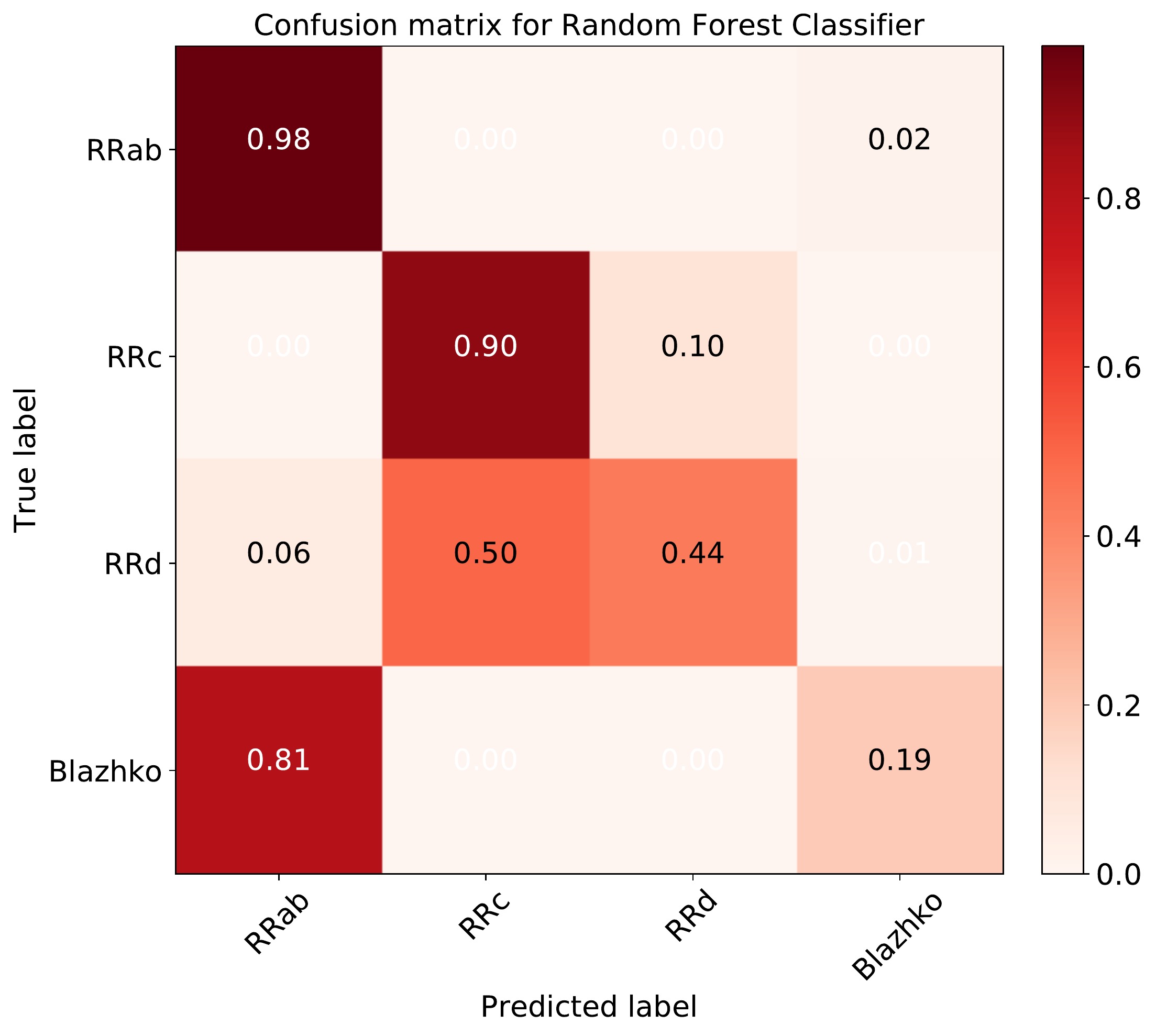}}
\subfloat[Third level]{\includegraphics[width=0.45\textwidth]{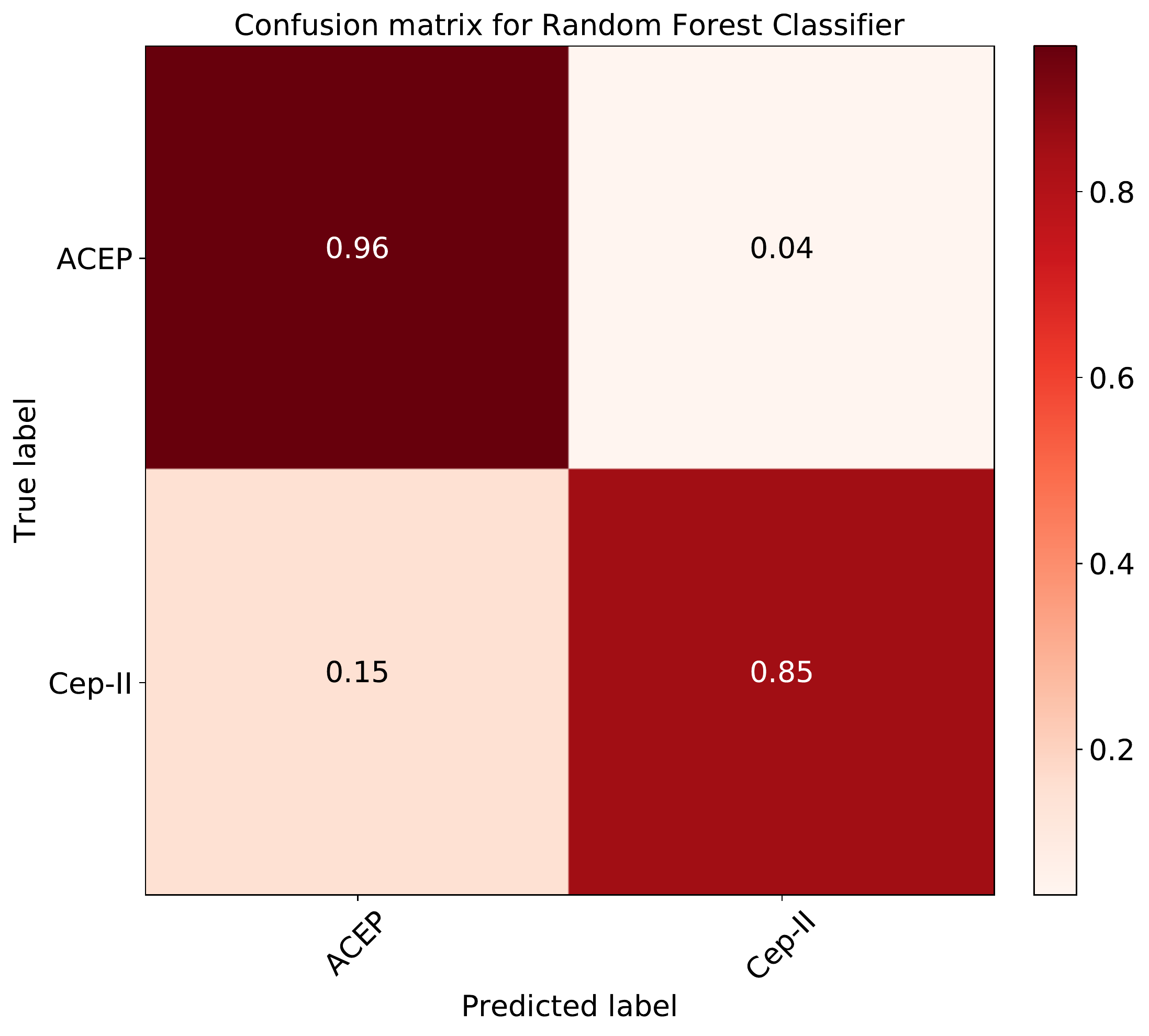}}
\caption{\label{fig:Confusion-matrices-for-Hierarchical-classification-classification}The normalized confusion matrices for the best classifier (RF) for hierarchical classification.}
\end{figure*}

Eventually, we follow the hierarchy down to the third layer where we investigate the categorization of two classes: RR Lyrae and Cepheids. Our aim here is to find to what extent we will be successful in distinguishing between the sub-classes of RR Lyrae (RRab, RRc, RRd and Blazhko) and Cepheids (ACEP and Cep-II). The analysis shows that we are successful in distinguishing between RRab and RRc with high balanced-accuracy. However, most of the RRd sources are classified as RRc and Blazhko sources are classified as RRab. To check whether our results are affected by class imbalance, we downsample RRc in the training set to the same number of samples as RRd and perform a binary classification. This process is repeated for RRab, whereby the number of objects is decreased to the same number as in Blazhko class. We train a binary classifier, keeping the test set the same for RRc \& RRd and RRab \& Blazhko. We found that using balanced classes does increase the performance of the classifier significantly, for instance, we are able to distinguish between RRab \& Blazhko and RRc \& RRd with a balanced-accuracy rate of 80 per cent and 75 per cent respectively.

\begin{figure*}
\centering
\subfloat[The period-amplitude distribution]{\includegraphics[width=0.45\textwidth]{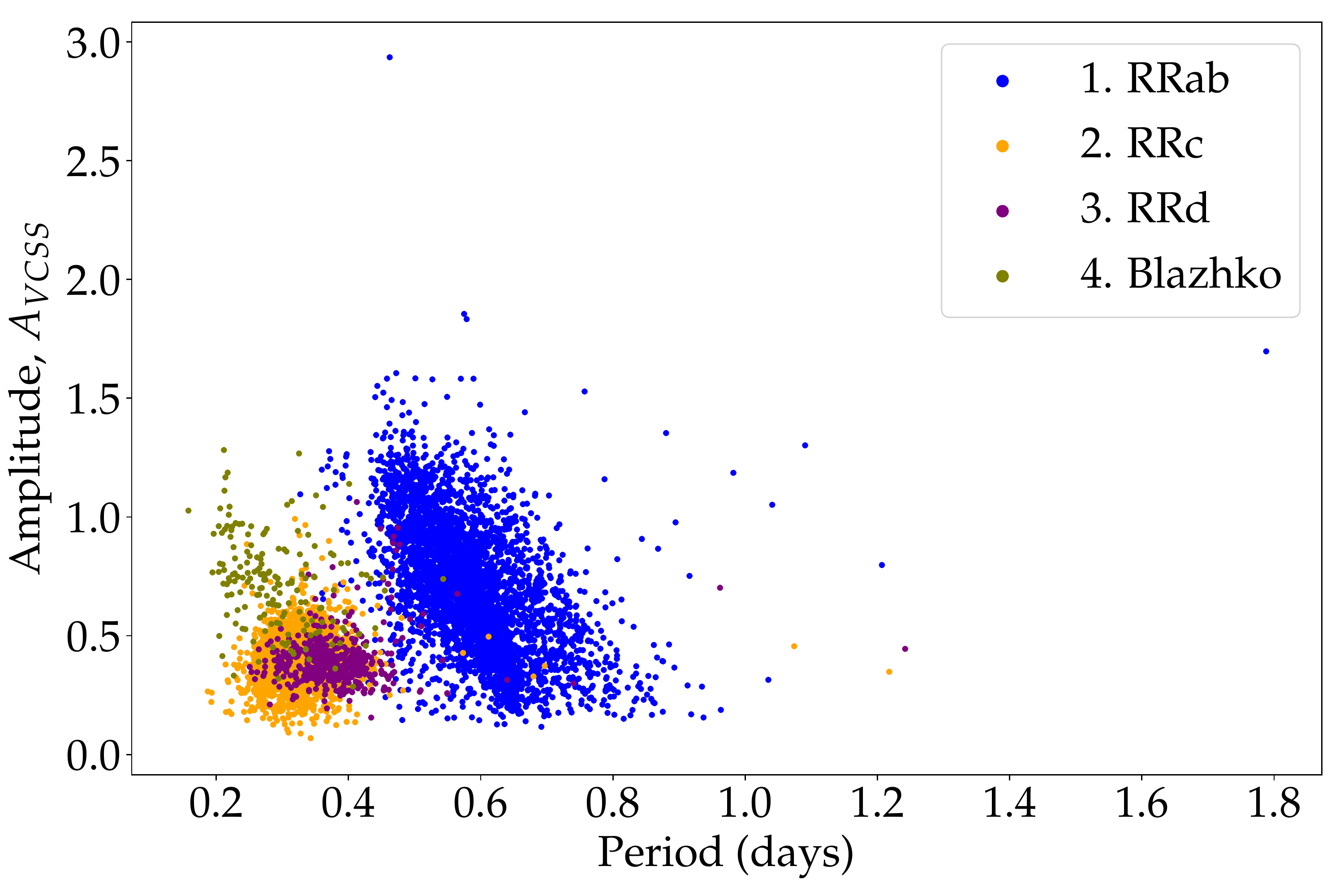}}
\subfloat[The period distribution]{\includegraphics[width=0.45\textwidth]{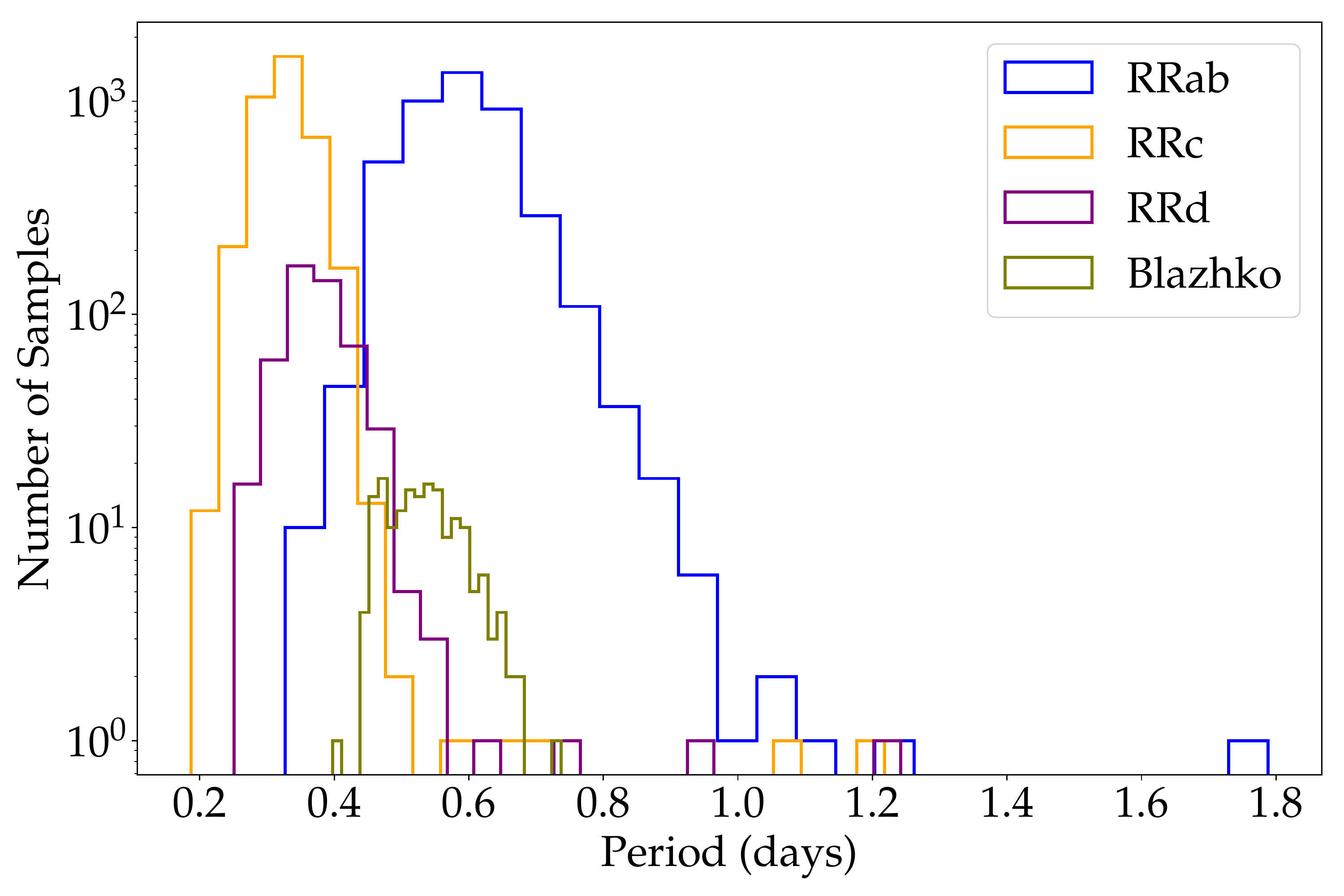}}
\caption{\label{fig:Analysis of RR Lyrae}We plot the distribution of RR Lyrae sub-types (RRab, RRc, RRd and Blazhko) using available data from the ascii catalog of the sources in \citet{Drake_2017}.}
\end{figure*}

For an in-depth analysis of the RR Lyrae sub-classification, we use the data provided by \citet{Drake_2017} that utilised the Adaptive Fourier Decomposition (AFD) method \citep{Torrealba_2015} to determine the period of each source. To see how the visually selected periodic variables differ from the initial candidates; we plot the amplitude and the period distribution of the variable stars available in the catalog. From Fig. \ref{fig:Analysis of RR Lyrae}, we note that it is a challenging task to separate the subclasses: RRab \& Blazhko and RRc \& RRd. We found that our ML classifier also struggles to separate those classes. In addition, we observe a clear separation between RRab and RRc, and our classification pipeline is also able to separate these two classes. In the same vein, \citet{Malz_2018} point out it is generally challenging to have a clear separation between RRc and RRd even though they have different pulsating modes and due to the rarity of RRd sources, they are often subsumed by RRc labels. Moreover, \citet{Drake_2017} argued that RRd phased light curves often appear like those of RRc and Blazhko stars often resemble RRab's when phase-folded over multiple cycles. 

Furthermore, we analyse the classification of Cepheids and we obtain an error classification rate of $\sim$19 per cent. On the same note, \citet{Drake_2017} argued that classifying ACEP and BL Her (a sub-groups of Cep-II) is difficult for field stars because of the mixture of stellar populations in the field and the inadequate distance information. 

For our hierarchical model, we present the RF classifier results only as it performs well compared to other classifiers discussed in this paper. From the results in Table \ref{tab:Hierarchical classification} and Fig. \ref{fig:Confusion-matrices-for-Hierarchical-classification-classification}, we see immediately a disparity in performance among the classes. This discrepancy in misclassification is due to the comparative size of each of the science classes. We note that we obtain high performance with classes that are data-rich. To alleviate the class-imbalance problem, one can either gather different catalogs for the under-sampled classes or augment the existing available catalogs via sophisticated statistical machine learning approaches.

\begin{figure}
\centering
\includegraphics[width=0.45\textwidth]{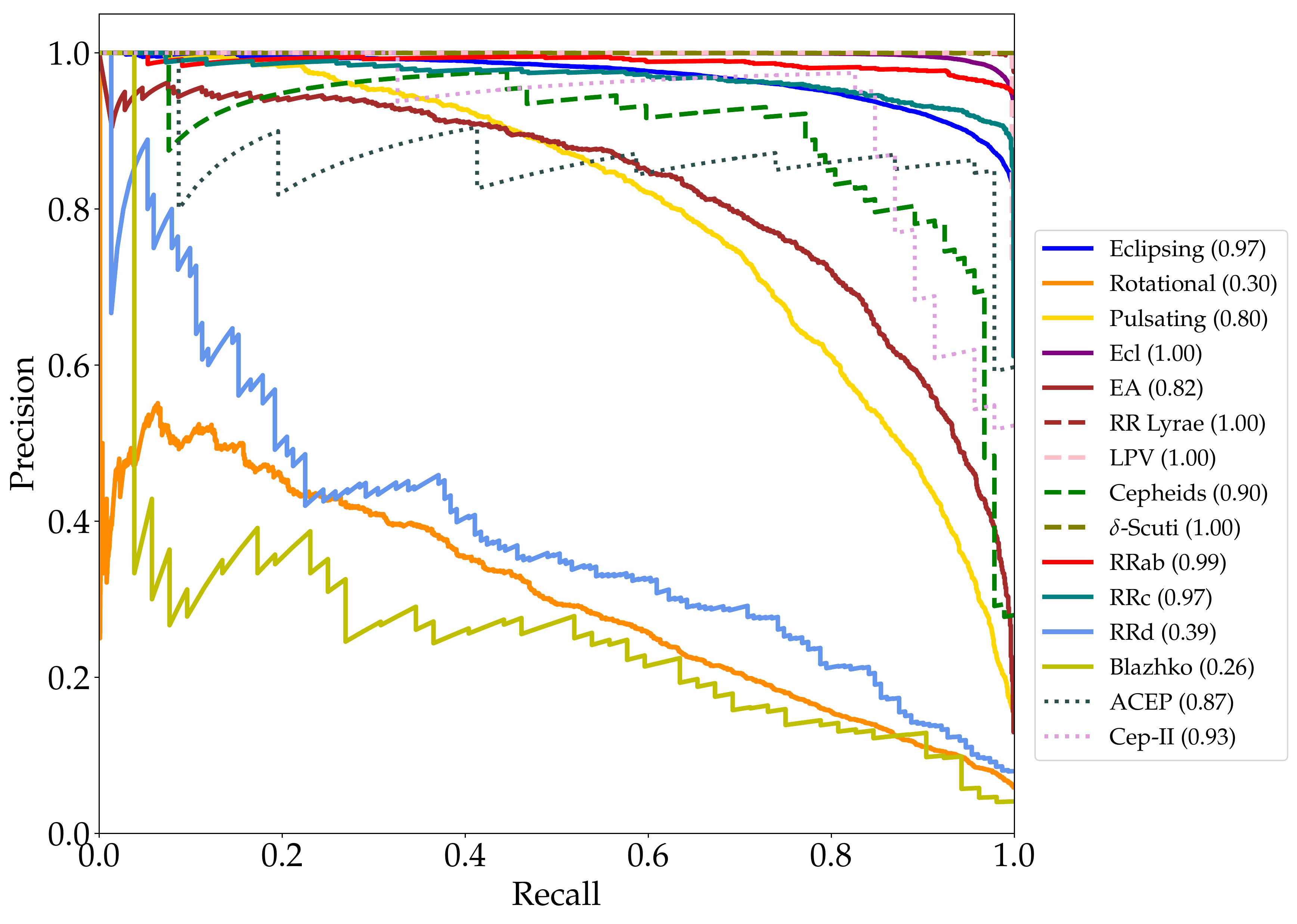}
\caption{\label{fig:learning_curves}Precision-Recall curves for each node in the hierarchical model. Each curve represents a different variable stars with the area under the precision-recall curves score in brackets. This metric is computed on the 30\% of the dataset used for testing, except that Type 5: Ecl stars has $\sim$15647 samples.}
\end{figure}

In Fig. \ref{fig:learning_curves}, we plot the precision-recall curve for each class and we note that the classification performance is very good. The area under the precision-recall curve values are greater than 0.85 for several classes, except for Type 7: Rotational, Type 3: RRd and Type 4: Blazhko. This correlates with the results presented in Table \ref{tab:Hierarchical classification}. In addition, we compare our proposed hierarchical approach to an already implemented machine learning package known as \texttt{UPSILON} \citep{kim2016package}. The latter used a RF classifier, trained on 16 features extracted from OGLE \citep{Udalski_1997} and EROS-2 \citep{Tisserand_2007} periodic variable stars light curves; while our model  is trained on 7 features from CRTS data. The comparison is made with only 8 classes out of 11 as \texttt{UPSILON} has not been trained on all the variable stars available in CRTS data. We report the results in terms of recall and F1-score in Table \ref{tab:Upsilon Results}. It is seen that our hierarchical model outperforms the \texttt{UPSILON} model in classifying most of the variable stars, except for Type 6: EA. We can plausibly say that our hierarchical classifier yields good performance with 7 features in classifying sub-groups of stars as compared to the \texttt{UPSILON} package. For a comparison in terms of features used, we report results when using a `flat' multi-class model (RF) in \S\ref{subsec:Multi-class classification} with 7 features and compare it with the hierarchical model. The hierarchical model outperforms both the \texttt{UPSILON} model and the multi-class model developed in this paper. This clearly shows that it is not necessary to extract many features to obtain higher classification metrics. In addition, we have shown that converting a `flat' multi-class problem into a hierarchical system improves variable star classification.

\begin{table*}
\begin{minipage}{150mm}
\caption{\label{tab:Upsilon Results}Comparison of our Hierarchical model, Multi-class model and \texttt{UPSILON} package \citep{kim2016package}. For a fair comparison, we test these models on 8 classes and report the scores in terms of recall and F1-score. The model with higher classification score in highlighted in blue.}
\noindent \begin{centering}
\begin{tabular}{c|cc|cc|cc}
\hline 
\multirow{3}{*}{\textbf{\textcolor{black}{Types of Variable Stars}} }& \multicolumn{2}{c|}{\textbf{\textcolor{black}{UPSILON Model}}} & \multicolumn{2}{c|}{\textbf{\textcolor{black}{Our Multi-Class Model}}} & \multicolumn{2}{c}{\textbf{\textcolor{black}{ Our Hierarchical Model}}}\tabularnewline

 & \multicolumn{2}{c|}{\textbf{\textcolor{black}{with 16 Features}}}  & \multicolumn{2}{c|}{\textbf{\textcolor{black}{with 7 Features}}}  & \multicolumn{2}{c}{\textbf{\textcolor{black}{with 7 Features}}} \tabularnewline
 
\cline{2-7} 
 & \textbf{\textcolor{black}{Recall }} & \textbf{\textcolor{black}{F1-Score}} & \textbf{\textcolor{black}{Recall }} & \textbf{\textcolor{black}{F1-Score}} & \textbf{\textcolor{black}{Recall}} & \textbf{\textcolor{black}{F1-Score}}\tabularnewline
\hline 
\hline
RRab & 0.75 & 0.86 & 0.92 & 0.73 & \textbf{\textcolor{blue}{0.98}} & \textbf{\textcolor{blue}{0.97}}\tabularnewline

RRc & 0.78 & 0.87 & 0.77 & 0.46 & \textbf{\textcolor{blue}{0.90}} & \textbf{\textcolor{blue}{0.91}}\tabularnewline

RRd & 0.11 & 0.20 & 0.33 & 0.14 & \textbf{\textcolor{blue}{0.44}} & \textbf{\textcolor{blue}{0.40}}\tabularnewline

Ecl (EC \& ESD) & 0.75 & 0.73 & 0.60 & 0.74 & \textbf{\textcolor{blue}{0.92}} & \textbf{\textcolor{blue}{0.95}}\tabularnewline

EA & \textbf{\textcolor{blue}{0.97}} &   \textbf{\textcolor{blue}{0.98}} &0.90 & 0.68 & 0.94 & 0.65\tabularnewline

LPV & 0.92 & 0.96 &0.98& 0.95 & \textbf{\textcolor{blue}{0.99}} & \textbf{\textcolor{blue}{1.00}}\tabularnewline

$\delta$-Scuti & 0.86 & 0.93 & 0.96 & 0.70 &\textbf{\textcolor{blue}{1.00}} & \textbf{\textcolor{blue}{0.98}}\tabularnewline
 
Cep-II & 0.38 & 0.55 & 0.52 & 0.32 &\textbf{\textcolor{blue}{0.85}} & \textbf{\textcolor{blue}{0.90}}\tabularnewline
\hline 
\end{tabular}
\par\end{centering}
\end{minipage}
\end{table*}

\section{Conclusion and Future work}\label{sec:conclusion}

With the upcoming synoptic surveys (for e.g. LSST, \citet{Ivezic_2008}), automated transient/variable-star classification is becoming an increasingly important field in astronomy. It presents a difficult computational challenge which we have nonetheless attempted to tackle in this work. We describe the core challenges associated with automatic variable-star classification and subsequently explored the nature of the data to be processed. We then applied various approaches with the aim of accurately classifying 11 types of variable star. Some of the methods we applied are similar to current techniques, while others are new to this field. Compared to other related work, we utilized only 7 input features during classification, where 6 features are based on statistical properties of the unfolded data and the Period, $T$, is obtained directly from the data catalog. We used these features as inputs to three separate commonly employed ML algorithms. We demonstrate that the RF classification algorithm performed best in all test cases. Treating the variable-star classification as a multi-class problem with 11 classes, results in a poor performance. In doing so, the RF  algorithm is efficient at identifying RRab, RRc, Ecl, EA, Rotational and LPV classes, but is unsuccessful in distinguishing, for example, $\delta$-Scuti and ACEP. However, by decomposing the multi-class problem into several binary classification problems, we gain a significant improvement in balanced-accuracy - with a balanced-accuracy rate of 1.0 for the classification of $\delta$-Scuti and ACEP. Our results suggest that decomposing a multi-class problem into several binary classification steps could yield improved results.

We therefore developed a hierarchical approach for classifying the 11-variable-star classes in our data, by aggregating those classes that show similarities. We show that a hierarchical taxonomy for n-class objects improves the classification rate. We are now successful in identifying sub-types of cepheids and eclipsing binaries with a balanced-accuracy rate of 81 per cent and 86 per cent respectively. Whilst the hierarchical approach does not work well for all classes, this is understandable in cases with high class-imbalances and a lack of training examples.  

We have presented an approach to analysing variable star data in such a way, that is beneficial to machine learning feature analysis and extraction. This process yields insights that allow us to obtain improved classification performance. In other words, we have taken a principled approach to feature analysis and design, especially for multi-class problems with a highly imbalanced datasets. We employ new methods for feature selection and evaluation and we show that converting a multi-class problem towards a hierarchical classification scheme helps to reduce the class-imbalance problem as well as provide a significant improvement for variable stars classification. In the near future, we wish to investigate the impact of data augmentation on the performance of our ML classifiers. Moreover, flagging data (as we did for  the ``Miscellaneous class") is often undesirable. One school of thought is to treat them as outliers but another might argue that these could perhaps indicate new discoveries. It is a major task with such a classification scheme, since we now have to tackle the problem in an unsupervised way, a topic currently in its infancy.

\section*{Acknowledgements}
We thank the referee for useful comments and suggestions in improving this manuscript. ZH acknowledges support from the UK Newton Fund as part of the Development in Africa with Radio Astronomy (DARA) Big Data project delivered via the Science \& Technology Facilities Council (STFC). BWS acknowledges funding from the European Research Council (ERC) under the European Union\textquotesingle s Horizon 2020 research and innovation programme (grant agreement No. 694745). AM is supported by the Imperial President's PhD Scholarship.







\appendix

\section{Evaluation Metrics}\label{sec:Evaluation_Metrics}
The performance of any machine learning algorithm is evaluated using measures such as the balanced-accuracy, the precision, the recall, the F1 score, sensitivity and specificity \citep{Chao}. They are defined in terms of True Positive $\left(T_{P}\right)$ Rate, False Positive $\left(F_{P}\right)$ Rate, True Negative $\left(T_{N}\right)$ Rate and False Negative $\left(F_{N}\right)$ Rate. The sensitivity metric is defined as the true positive rate or positive class accuracy, while specificity is referred to as the true negative rate or equivalently negative class accuracy \citep{Danjuma2015PerformanceEO}.
\begin{itemize}

\item Sensitivity Measure: Equation \ref{eq:Recall} also known as the
true positive rate or ``recall". It measures the proportion of actual
positives correctly identified by the model.
\end{itemize}
\begin{equation}
\textrm{Sensitivity/Recall}\,=\,\frac{T_{P}}{T_{P}+F_{N}}\label{eq:Recall}
\end{equation}

\begin{itemize}
\item Specificity Measure: Equation \ref{eq:Specificity} also known as
True Negative rate. It measures how well a model identifies negative
results.
\end{itemize}
\begin{equation}
\textrm{Specificity}\,=\,\frac{T_{N}}{T_{N}+F_{P}}\label{eq:Specificity}
\end{equation}

\begin{itemize}
\item Precision: It is a measure of retrieved instances that are correctly labelled. Precision is described in Equation \ref{eq:Precision}.
\end{itemize}
\begin{equation}
\textrm{Precision}\,=\,\frac{T_{P}}{T_{P}+F_{P}}\label{eq:Precision}
\end{equation}

\begin{itemize}
\item F1-score: It is a metrics that aims to quantify overall performance, expressed in terms of precision and recall as shown in Equation \ref{eq:F1-score}.
\end{itemize}
\begin{equation}
\textrm{F1-Score}\,=\,\frac{2\times\textrm{Precision}\times\textrm{Recall}}{\textrm{Precision}+\textrm{Recall}}\label{eq:F1-score}
\end{equation}

\begin{itemize}
\item Balanced accuracy measure: It is defined as the average of recall obtained on each class and it is a metric that deal with imbalanced classes.
\end{itemize}
\begin{equation}
\textrm{Balanced Accuracy}\,=\,\frac{\textrm{Sensitivity}+\textrm{Specificity}}{2}\times100\%\label{eq:Accuracy}
\end{equation}

\begin{itemize}
\item Geometric Mean Score: G-Mean is defined as the squared root of the product of class-wise sensitivity. It maximizes the accuracy on each class in addition to keep these accuracies balanced.
\end{itemize}
\begin{equation}
\textrm{G-Mean}\,=\,\sqrt{\textrm{Sensitivity} \times \textrm{Specificity}}\label{eq:G-Mean}
\end{equation}

\begin{itemize}
\item Precision-Recall curve: PR curve illustrates the trade-off between TPR and positive predictive value for a model at different probability thresholds.
\end{itemize}

\begin{itemize}
\item Confusion Matrices: The $T_{P},\,F_{P},\,T_{N},\,\textrm{and\,}F_{N}$ can be visualized by a confusion matrix as illustrated in Table \ref{tab:Confusion-matrix-theory}, where the predicted class is indicated in each column and the actual class in each row. In this case, from Table \ref{tab:Confusion-matrix-theory}, the true positives $\left(T_{P}\right)$ are Class I examples that were correctly classified as Class I, false positives $F_{P}$ correspond to Class II examples wrongly classified as Class I. In a similar way, false negatives $F_{N}$ and true negatives $T_{N}$ can be explained. For binary classification problems, Class I corresponds to positive class and Class II corresponds to negative class. After the training and testing process, we evaluate our pipeline using balanced-accuracy, G-mean, F1-score, recall values and confusion matrices.
\end{itemize}

\begin{table}
\caption[Confusion matrix]{Confusion matrix implemented for our specific problem, where the
positive class corresponds to Class I and the negative class to Class
II for the two classification schemes. True/false positives/negatives
are represented as $T_{P},\,F_{P},\,T_{N},\,\textrm{and\,}F_{N}$
respectively.\label{tab:Confusion-matrix-theory}}

\renewcommand{\arraystretch}{1.5}
\centering{}%
\begin{tabular}{>{\raggedright}m{1.5cm}ccc}
 &  & Class II & Class I\tabularnewline
\hline 
\multirow{2}{1.5cm}{\textbf{Actual Class}} & Class II & $T_{N}$ & $F_{P}$\tabularnewline
\cline{2-4} 
 & Class I & $F_{N}$ & $T_{P}$\tabularnewline
\hline 
 &  & \multicolumn{2}{c}{\textbf{Predicted Class}}\tabularnewline
\end{tabular}
\end{table}

\bibliographystyle{mnras}
\bibliography{reference.bib}

\begin{thebibliography}{}
\makeatletter
\relax
\def\mn@urlcharsother{\let\do\@makeother \do\$\do\&\do\#\do\^\do\_\do\%\do\~}
\def\mn@doi{\begingroup\mn@urlcharsother \@ifnextchar [ {\mn@doi@}
  {\mn@doi@[]}}
\def\mn@doi@[#1]#2{\def\@tempa{#1}\ifx\@tempa\@empty \href
  {http://dx.doi.org/#2} {doi:#2}\else \href {http://dx.doi.org/#2} {#1}\fi
  \endgroup}
\def\mn@eprint#1#2{\mn@eprint@#1:#2::\@nil}
\def\mn@eprint@arXiv#1{\href {http://arxiv.org/abs/#1} {{\tt arXiv:#1}}}
\def\mn@eprint@dblp#1{\href {http://dblp.uni-trier.de/rec/bibtex/#1.xml}
  {dblp:#1}}
\def\mn@eprint@#1:#2:#3:#4\@nil{\def\@tempa {#1}\def\@tempb {#2}\def\@tempc
  {#3}\ifx \@tempc \@empty \let \@tempc \@tempb \let \@tempb \@tempa \fi \ifx
  \@tempb \@empty \def\@tempb {arXiv}\fi \@ifundefined
  {mn@eprint@\@tempb}{\@tempb:\@tempc}{\expandafter \expandafter \csname
  mn@eprint@\@tempb\endcsname \expandafter{\@tempc}}}

\bibitem[\protect\citeauthoryear{Bates, Bailes, Bhat, Burgay  et~al.}{Bates
  et~al.}{2011}]{Bates_2011a}
Bates S.,  Bailes M.,  Bhat N.,  Burgay M.,   et~al., 2011, Monthly Notices of
  the Royal Astronomical Society, 416, 2455

\bibitem[\protect\citeauthoryear{Belokurov, Evans  \& Le~Du}{Belokurov
  et~al.}{2003}]{Belokurov2003}
Belokurov V.,  Evans N.~W.,   Le~Du Y.,  2003, Monthly Notices of the Royal
  Astronomical Society, 341(4), 1373

\bibitem[\protect\citeauthoryear{Bentley}{Bentley}{1975}]{Bentley_1975}
Bentley J.~L.,  1975, Communications of the ACM, 18 (9), 509

\bibitem[\protect\citeauthoryear{Bergstra, Yamins  \& Cox}{Bergstra
  et~al.}{2013}]{bergstra2013hyperopt}
Bergstra J.,  Yamins D.,   Cox D.~D.,  2013, in Proceedings of the 12th Python
  in science conference. pp 13--20

\bibitem[\protect\citeauthoryear{Blazhko}{Blazhko}{1907}]{Blazhko_1907}
Blazhko S.,  1907, Astronomische Nachrichten, 175, 325

\bibitem[\protect\citeauthoryear{Breiman}{Breiman}{2001}]{Breiman_2001}
Breiman L.,  2001, Machine Learning, 45, 5

\bibitem[\protect\citeauthoryear{Brown, Pocock, Zhao  \& Luj{\'a}n}{Brown
  et~al.}{2012}]{brown2012conditional}
Brown G.,  Pocock A.,  Zhao M.-J.,   Luj{\'a}n M.,  2012, Journal of machine
  learning research, 13, 27

\bibitem[\protect\citeauthoryear{Buturovic}{Buturovic}{1993}]{Buturovic_1993}
Buturovic L.~J.,  1993, Pattern Recognition, 26, 611

\bibitem[\protect\citeauthoryear{Catelan \& Smith}{Catelan \&
  Smith}{2015}]{Catelan_2015}
Catelan M.,  Smith H.~A.,  2015, Wiley-VCH

\bibitem[\protect\citeauthoryear{Cauchy}{Cauchy}{1853}]{Cauchy_1853}
Cauchy A.,  1853, C.R. Acad. Sci, 37, 198

\bibitem[\protect\citeauthoryear{Chao, Liaw  \& Breiman}{Chao
  et~al.}{2004}]{Chao}
Chao C.,  Liaw A.,   Breiman L.,  2004, University of California, Berkeley

\bibitem[\protect\citeauthoryear{Danjuma}{Danjuma}{2015}]{Danjuma2015PerformanceEO}
Danjuma K.~J.,  2015, CoRR, abs/1504.04646

\bibitem[\protect\citeauthoryear{Dietterich}{Dietterich}{2000}]{Dietterich_2000}
Dietterich T.~G.,  2000, Multiple Classifier Systems, 1857, 1

\bibitem[\protect\citeauthoryear{Djorgovski, Donalek, Mahabal
  et~al.}{Djorgovski et~al.}{2011}]{Djorgovski2011}
Djorgovski S.~G.,  Donalek C.,  Mahabal A.,   et~al., 2011, preprint

\bibitem[\protect\citeauthoryear{Djorgovski, Graham, Donalek
  et~al.}{Djorgovski et~al.}{2016}]{Djorgovski2016}
Djorgovski S.~G.,  Graham M.~J.,  Donalek C.,   et~al., 2016, preprint

\bibitem[\protect\citeauthoryear{Drake, Djorgovski, Mahabal  et~al.}{Drake
  et~al.}{2009}]{Drakke2009}
Drake A.~J.,  Djorgovski S.~G.,  Mahabal A.,   et~al., 2009, ApJ, 696, 870

\bibitem[\protect\citeauthoryear{Drake, Djorgovski, Catelan, Graham
  et~al.}{Drake et~al.}{2017}]{Drake_2017}
Drake A.~J.,  Djorgovski S.~G.,  Catelan M.,  Graham M.~J.,   et~al., 2017,
  Monthly Notices of the Royal Astronomical Society, 469 (3), 3688

\bibitem[\protect\citeauthoryear{Eyer \& Blake}{Eyer \&
  Blake}{2005}]{Eyer_2005}
Eyer L.,  Blake C.,  2005, MNRAS

\bibitem[\protect\citeauthoryear{Gregory \& Loredo}{Gregory \&
  Loredo}{1992}]{Gregory_1992}
Gregory P.~C.,  Loredo T.~J.,  1992, Astrophysical Journal

\bibitem[\protect\citeauthoryear{Gupta}{Gupta}{1960}]{Gupta_1960}
Gupta S.,  1960, Psychometrika, 25(4), 393

\bibitem[\protect\citeauthoryear{Guyon \& Elisseeff}{Guyon \&
  Elisseeff}{2003}]{Guyon_2003}
Guyon I.,  Elisseeff A.,  2003, Journal of Machine Learning Research, 3, 1157

\bibitem[\protect\citeauthoryear{He \& Garcia}{He \&
  Garcia}{2008}]{he2008learning}
He H.,  Garcia E.~A.,  2008, IEEE Transactions on Knowledge \& Data
  Engineering, pp 1263--1284

\bibitem[\protect\citeauthoryear{Ivezic, Kahn, Tyson  et~al.}{Ivezic
  et~al.}{2008}]{Ivezic_2008}
Ivezic Z.,  Kahn S.~M.,  Tyson J.~A.,   et~al., 2008, American Astronomical
  Society

\bibitem[\protect\citeauthoryear{Juric, Kantor, Lim, Lupton  et~al.}{Juric
  et~al.}{2015}]{Juric_2015}
Juric M.,  Kantor J.,  Lim K.~T.,  Lupton R.~H.,   et~al., 2015, Astronomical
  Data Analysis Software and Systems, 512

\bibitem[\protect\citeauthoryear{Kim \& Bailer-Jones}{Kim \&
  Bailer-Jones}{2016}]{kim2016package}
Kim D.-W.,  Bailer-Jones C.~A.,  2016, Astronomy \& Astrophysics, 587, A18

\bibitem[\protect\citeauthoryear{Kullback \& Leibler}{Kullback \&
  Leibler}{1951}]{Kullback_1951}
Kullback S.,  Leibler R.~A.,  1951, The Annals of Mathematical Statistics,
  22(1), 79

\bibitem[\protect\citeauthoryear{Last, Douzas  \& Bacao}{Last
  et~al.}{2017}]{Last_2017}
Last F.,  Douzas G.,   Bacao F.,  2017, arXiv:1711.00837

\bibitem[\protect\citeauthoryear{Lochner, McEwen, Peiris  et~al.}{Lochner
  et~al.}{2016}]{Lochner_2016}
Lochner M.,  McEwen J.~D.,  Peiris H.~V.,   et~al., 2016, The Astrophysical
  Journal Supplement Series, 225(1), 14

\bibitem[\protect\citeauthoryear{Lomb}{Lomb}{1976}]{Lomb_1976}
Lomb N.~R.,  1976, Astrophysics and Space Science, 39, 447

\bibitem[\protect\citeauthoryear{Lyon, Stappers, Cooper, Brooke  \&
  Knowles}{Lyon et~al.}{2016}]{Lyon_2016}
Lyon R.~J.,  Stappers B.~W.,  Cooper S.,  Brooke J.~M.,   Knowles J.~D.,  2016,
  Monthly Notices of the Royal Astronomical Society, 459, 1104

\bibitem[\protect\citeauthoryear{Mahabal, Donalek, Djorgovski  et~al.}{Mahabal
  et~al.}{2012}]{Mahabal2012}
Mahabal A.~A.,  Donalek C.,  Djorgovski S.~G.,   et~al., 2012, New Horizons in
  Time Domain Astronomy, 285, 355

\bibitem[\protect\citeauthoryear{Mahabal, Sheth, Gieseke  et~al.}{Mahabal
  et~al.}{2017}]{Mahabal2017}
Mahabal A.,  Sheth K.,  Gieseke F.,   et~al., 2017, IEEE Symposium Series on
  Computational Intelligence, p.~2757

\bibitem[\protect\citeauthoryear{Malz, Hlozek, Allam, Bahmanyar  et~al.}{Malz
  et~al.}{2018}]{Malz_2018}
Malz A.,  Hlozek R.,  Allam T.~J.,  Bahmanyar A.,   et~al., 2018,
  arXiv:1809.11145

\bibitem[\protect\citeauthoryear{Narayan, Zaidi, Soraisam  et~al.}{Narayan
  et~al.}{2018}]{Narayan_2018}
Narayan G.,  Zaidi T.,  Soraisam M.~D.,   et~al., 2018, Astrophysical Journal
  Supplement Series, 236(1)

\bibitem[\protect\citeauthoryear{Nun, Protopapas, Sim  et~al.}{Nun
  et~al.}{2015}]{Nun_2015}
Nun I.,  Protopapas P.,  Sim B.,   et~al., 2015, arXiv:1506.00010

\bibitem[\protect\citeauthoryear{Pearson}{Pearson}{1895}]{Pearson_1895}
Pearson K.,  1895, Proceedings of the Royal Society of London, 58, 240

\bibitem[\protect\citeauthoryear{Pedregosa et~al.}{Pedregosa
  et~al.}{2011}]{Pedregosa}
Pedregosa F.,  et~al., 2011, Journal of Machine Learning Research, 12, 2825

\bibitem[\protect\citeauthoryear{Quinlan}{Quinlan}{1986}]{Quinlan_1986}
Quinlan J.~R.,  1986, Machine Learning, 1, 81

\bibitem[\protect\citeauthoryear{Revsbech, Trotta  \& van Dyk}{Revsbech
  et~al.}{2018}]{Revsbech_2018}
Revsbech E.~A.,  Trotta R.,   van Dyk D.~A.,  2018, MNRAS, 473, 3969

\bibitem[\protect\citeauthoryear{Richards et~al.,}{Richards
  et~al.}{2011}]{richards2011machine}
Richards J.~W.,  et~al., 2011, The Astrophysical Journal, 733, 10

\bibitem[\protect\citeauthoryear{Saha \& Vivas}{Saha \&
  Vivas}{2017}]{Saha_2017}
Saha A.,  Vivas A.~K.,  2017, The Astronomical Journal, 154

\bibitem[\protect\citeauthoryear{Scargle}{Scargle}{1982}]{Scargle_1982}
Scargle J.~D.,  1982, Astrophysical Journal, 263, 835

\bibitem[\protect\citeauthoryear{Silla \& Freitas}{Silla \&
  Freitas}{2011}]{Silla_2011}
Silla C. N.~J.,  Freitas A.~A.,  2011, Data Mining and Knowledge Discovery, 22,
  31

\bibitem[\protect\citeauthoryear{{Tisserand} et~al.,}{{Tisserand}
  et~al.}{2007}]{Tisserand_2007}
{Tisserand} P.,  et~al., 2007, \mn@doi [\aap] {10.1051/0004-6361:20066017},
  \href {https://ui.adsabs.harvard.edu/abs/2007A&A...469..387T} {469, 387}

\bibitem[\protect\citeauthoryear{Torrealba, Catelan, Drake, Djorgovski
  et~al.}{Torrealba et~al.}{2015}]{Torrealba_2015}
Torrealba G.,  Catelan M.,  Drake A.~J.,  Djorgovski S.~G.,   et~al., 2015,
  Monthly Notices of the Royal Astronomical Society, 446, 2251

\bibitem[\protect\citeauthoryear{{Udalski}, {Kubiak}  \& {Szymanski}}{{Udalski}
  et~al.}{1997}]{Udalski_1997}
{Udalski} A.,  {Kubiak} M.,   {Szymanski} M.,  1997, \actaa, \href
  {https://ui.adsabs.harvard.edu/abs/1997AcA....47..319U} {47, 319}

\bibitem[\protect\citeauthoryear{Wattenberg, Fernanda  \& Johnson}{Wattenberg
  et~al.}{2016}]{wattenberg2016how}
Wattenberg M.,  Fernanda V.,   Johnson I.,  2016, \mn@doi [Distill]
  {10.23915/distill.00002}

\bibitem[\protect\citeauthoryear{Willemsen \& Eyer}{Willemsen \&
  Eyer}{2007}]{Willemsen_2007}
Willemsen P.,  Eyer L.,  2007, arXiv:0712.2898v1

\bibitem[\protect\citeauthoryear{Yang \& Fong}{Yang \& Fong}{2011}]{Yang_2011}
Yang H.,  Fong S.,  2011, Proceedings of the 13th international conference on
  Data warehousing and knowledge discovery, pp 471--483

\bibitem[\protect\citeauthoryear{Zhang}{Zhang}{2004}]{Zhang_2004}
Zhang H.,  2004, FLAIR, 2

\bibitem[\protect\citeauthoryear{van~der Maaten \& Hinton}{van~der Maaten \&
  Hinton}{2008}]{vanderMaaten_2008}
van~der Maaten L.,  Hinton G.,  2008, Journal of Machine Learning Research, pp
  2579--2605

\makeatother
\end{thebibliography}

\bsp	
\label{lastpage}
\end{document}